\documentclass[aps,prx,reprint]{revtex4-2}

\usepackage{amsmath,amssymb,amsfonts}
\usepackage{graphicx}
\usepackage{epsfig}
\usepackage{color}
\usepackage{esint}

\usepackage{lipsum} 
\usepackage{printlen}

\DeclareUnicodeCharacter{3B1}{\ensuremath{\alpha}}
\DeclareUnicodeCharacter{3B2}{\ensuremath{\beta}}

\begin{document}
	
	
	\title{Spatio-Temporal Dynamics of Nucleo-Cytoplasmic Transport}
	
	
	\author{S.\ Alex Rautu$^1$}
	
	\author{Alexandra Zidovska$^3$}
	
	\author{Michael J.\ Shelley$^{1,2}$}
	\email{mshelley@flatironinstitute.org}
	\affiliation{
		$^1$Center for Computational Biology, Flatiron Institute, New York, NY 10010, USA\\
		$^2$Courant Institute, New York University, New York, NY 10012, USA\\
		$^3$Center for Soft Matter Research, Department of Physics, New York University, New York, NY 10003, USA
	}
	
	\begin{abstract}
		Nucleocytoplasmic transport is essential for cellular function, presenting a canonical example of rapid molecular sorting inside cells. It consists of a coordinated interplay between import/export of molecules in/out the cell nucleus. Here, we investigate the role of spatio-temporal dynamics of the nucleocytoplasmic transport and its regulation. We develop a biophysical model that captures the main features of the nucleocytoplasmic transport, in particular, its regulation through the Ran cycle. Our model yields steady-state profiles for the molecular components of the Ran cycle, their relaxation times, as well as the nuclear-to-cytoplasmic molecule ratio. We show that these quantities are affected by their spatial dynamics and heterogeneity within the nucleus. Specifically, we find that the spatial nonuniformity of Ran Guanine Exchange Factor (RanGEF)---particularly its proximity to the nuclear envelope--- increases the Ran content in the nucleus. We further show that RanGEF's accumulation near the nuclear envelope results from its intrinsic dynamics as a nuclear cargo, transported by the Ran cycle itself. Overall, our work highlights the critical role of molecular spatial dynamics in cellular processes, and proposes new avenues for theoretical and experimental inquiries into the nucleocytoplasmic transport.
		
	\end{abstract}
	
	
	
	
	\maketitle
	
	\section{Introduction}
	
	Eukaryotic cells are organized into functional compartments, such as the nucleus and the cytoplasm. The former houses the genome and DNA-related machinery, the latter contains cytosol and organelles for cellular functions such as protein synthesis and degradation \cite{Alberts2008}. The boundary between the nucleus and the cytoplasm is delineated by the nuclear envelope (NE) \cite{Alberts2008}, with molecules transported bidirectionally across the NE through nuclear pore complexes (NPCs) \cite{Wente2010}. Small molecules, such as ions and nucleotides, can freely diffuse through the NPCs \cite{Timney2016,Keminer1999,Mohr2009}, while larger molecules, such as proteins, require a regulated multi-step process~\cite{Benarroch2019, Macara2001}. These large cargoes bind to a nuclear transport receptor (NTR), which then transports them in/out of the nucleus, depending if the cargo has a nuclear localization or export signal (NLS/NLS)~\cite{Macara2001,Cautain2015}.\\
	\indent	
	While individual molecular translocations through NPCs do not require energy, being thermally driven and facilitated by interactions with nucleoporins inside the NPC~\cite{Zilman2017, Zilman2023, Yang2004, Ribbeck1999}, they are part of a complex cycle that is essentially an energy-driven pump~\cite{Ribbeck1999, Gorlich1999, Cavazza2016, Kubitscheck2005, Hoogenboom2021}. This cycle can generate import/export fluxes against concentration gradients, maintaining the system in a non-equilibrium steady state~\cite{Hoogenboom2021}. This import/export cycle is run by GTP and the asymmetric distribution of a GTPase Ran protein across the NE~\cite{Joseph2006}, with RanGDP largely in the cytoplasm and RanGTP in the nucleus~\cite{Yang2004,Ribbeck1999,Gorlich1999,Cavazza2016}. This asymmetry is established by the guanine nucleotide exchange factor (RanGEF) \cite{Li2003,Hutchins2004,Cushman2004,Hitakomate2010}, which promotes GDP-to-GTP exchange in the nucleus, as well as the GTPase-activating protein (RanGAP) \cite{Seewald2002}, which in turn facilitates GTP hydrolysis in the cytoplasm. Importantly, RanGEF is bound to chromatin, while RanGAP associates with the cytoplasmic side of NPCs.\\
	\indent
	The import--export cycles of cargo proteins are tightly regulated by RanGTP. Figure 1 depicts the import cycle, with each cycle using one GTP molecule and resulting in the export of one Ran molecule per cargo \footnote{For transport proteins like Importin-$\beta$, which utilize an adaptor protein, Importin-$\alpha$, to attach to the cargo, the energy expenditure involves two molecules of GTP.}. Taken together, the import--export cycles are tightly controlled by the Ran cycle ~\cite{Hoogenboom2021}. Indeed, reversing the RanGTP gradient between the nucleus and cytoplasm leads to an inverted gradient of the cargoes~\cite{Nachury1999}.\\
	\indent
	To date, this system has been primarily modeled using kinetic rate equations \cite{Wang2017, Gorlich2003, Smith2002a,Kopito2007,Kopito2009,Timney2006,Ribbeck2001, Riddick2007, Kim2013, Kim2013b}, while assuming a rapid homogenization of the Ran cycle components within the nucleus and cytoplasm~\cite{Wu2009, Abu-Arish2009a,Timney2016, Ribbeck2001}. However, this assumption breaks down if the sources, or sinks, of the molecular components exhibit a spatial structure. Importantly, the nuclear RanGTP localization depends on the localization of RanGEF \cite{Li2003,Hutchins2004,Cushman2004,Hitakomate2010}. Thus, a heterogeneous RanGEF distribution in the nucleus would lead to a non-uniform distribution of RanGTP. Indeed, the localization of RanGEF regulates the RanGTP gradient, highlighting the role of spatial dynamics in nuclear transport. Notably, RanGEF shows a higher affinity for heterochromatin \cite{Paschal2011,Paschal2014,Paschal2015,Paschal2019, Casolari2004}, which is predominantly located at the nuclear periphery and nucleoli \cite{Bizhanova2021,Zidovska2020}. Hence, RanGTP concentration may be higher near the NE~\cite{Abu-Arish2009a}. Also, RanGTP is found to aid the dissociation of some NTRs from the NPCs \cite{Liphardt2015}, which would require a steady RanGTP supply at the NE. These intricate spatial effects are not accounted for by a spatially homogeneous model.\\
	\indent
	In this study, we investigate the mechanisms behind the nucleo-cytoplasmic transport mediated by the Ran cycle. In contrast to previous studies, which assumed a rapid homogenization of the Ran cycle components in the nucleus and cytoplasm~\cite{Wang2017, Gorlich2003, Smith2002a,Kopito2007,Kopito2009,Timney2006, Riddick2007, Kim2013, Kim2013b, Wu2009, Abu-Arish2009a,Timney2016, Ribbeck2001}, we explore the role of heterogeneity and spatial dynamics of the Ran cycle. To this end, we develop a model of the spatiotemporal transport of Ran components and explore how spatial heterogeneity of RanGEF affects the Ran cycle. We find analytical solutions for steady-state concentrations and relaxation times of Ran components, and obtain ratios of the nuclear to cytoplasmic Ran. We find that localization of RanGEF near the NE significantly increases the Ran content in the nucleus. Lastly, we show that a heterogeneous distribution of RanGEF near the NE originates from its dynamics as a NLS-containing cargo. 
	\color{black} 

	\begin{figure}[t!]\includegraphics[width=\columnwidth]{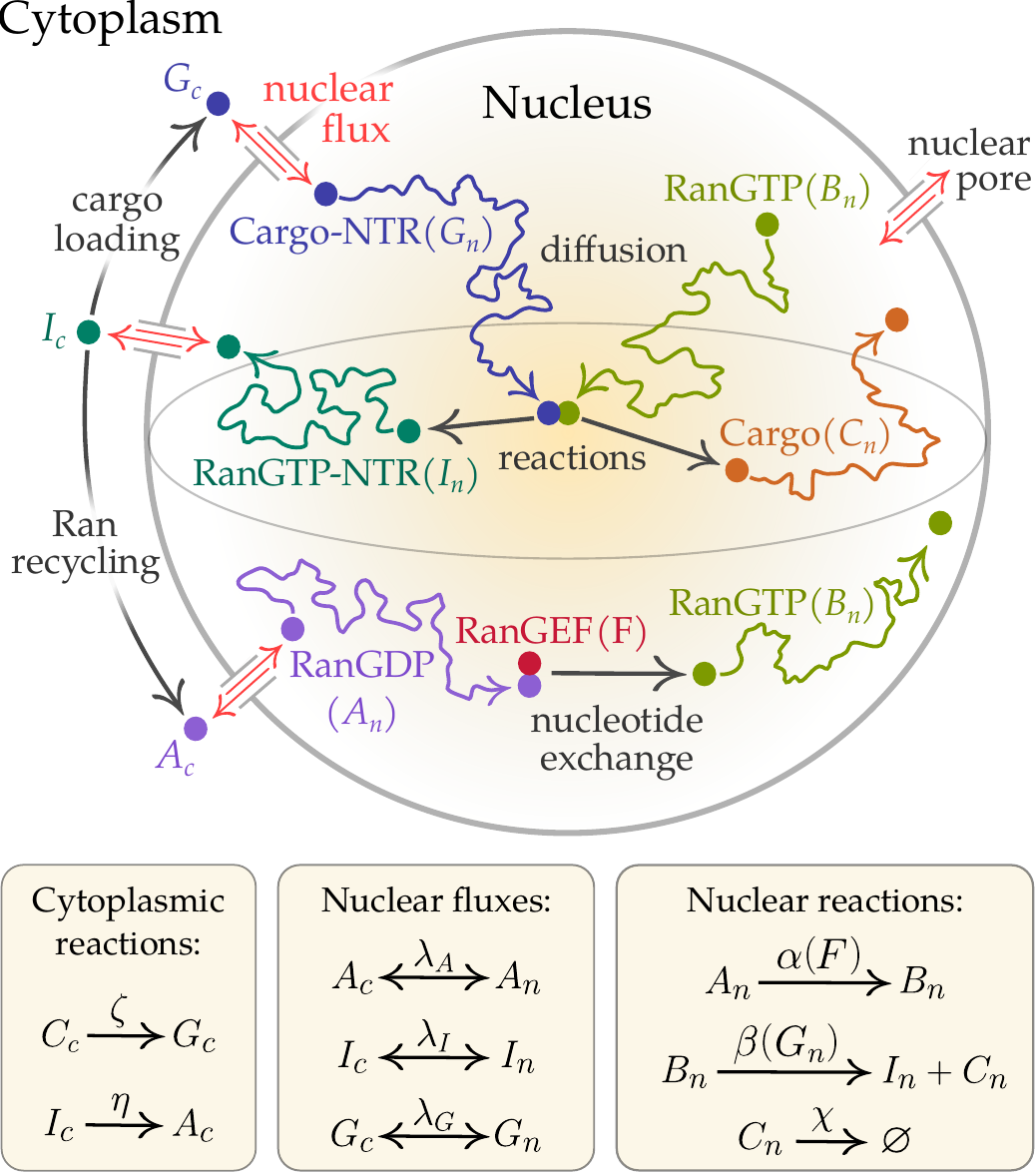}
	\caption{\label{fig:figure_1} 
		Ran-mediated nuclear transport of a cargo molecule, forming a complex with a nuclear transport receptor (NTR). Cargo-NTR complex passes through nuclear pores, diffuses, and releases the cargo upon RanGTP binding. The RanGTP--NTR complex is recycled to the cytoplasm, where it is dissociated through GTP hydrolysis. RanGDP reenters the nucleus and converts to RanGTP via RanGEF, completing its cycle. {The boxes tabulate the reaction rates in our model for each corresponding domain. In the cytoplasm, $\zeta$ denotes the rate of cargo loading onto NTRs, and $\eta$ is the hydrolysis rate of RanGTP. Nuclear pore permeability is denoted by $\lambda$. In the nucleus, the Ran nucleotide exchange is given by local rate $\alpha$, the unloading rate of cargoes from NRT complexes is denoted by $\beta$, and $\chi$ is the cargo depletion rate.}}
	\end{figure}
	
	\section{Model}
 
	We model the nucleus as a sphere of radius $R$ with boundary $\mathcal{S}$ (the NE; $R\simeq5\,\text{--}\,10\;\mu\mathrm{m}$~\cite{Alberts2008}) enclosing a volume~$\mathcal{V}_n$. We assume a constant NPC surface density, providing a uniform flux of particles across the NE \cite{Winey1997}. Each NPC supports $10^2$--$10^3$ translocations per second \cite{Ribbeck2001}. The outwards flux at the NE of a molecular species $X$ can be described by $\mathcal{J}_X=\lambda_X\left(X_{c}-X_{n}\right)$~\cite{Wang2017}, where $\lambda_X$ is the permeability, and the subscripts ${c}$ and ${n}$ denote the cytoplasm and nucleus, respectively. Here, the concentration is the number of molecules per unit volume, ignoring the presence of subnuclear bodies and the polymeric structure of chromatin. The nuclear molecules are assumed to diffuse in the nucleus with an effective constant diffusivity, denoted by $d_X$ for a solute $X$, in the range $1-20\,\mu\text{m}^2$/s \cite{Wu2009, Abu-Arish2009a}. We do not explicitly account for advection of molecules by nucleoplasmic flows \cite{Ashwin2019, Mine-Hattab2020, Oliveira2021, Oshidari2020a, Barth2020,  Zidovska2013, Saintillan2018, Zidovska2020, Saintillan2022}, with such an effect, if present, entering only as a contribution to the apparent diffusion constant. We assume that cytoplasmic concentrations are uniform, akin to the experimental conditions of permeabilized cells where the external delivery of molecules is controlled~\cite{Yang2004}.  Thus, at the NE the diffusive flux of molecules is given by
	\begin{equation}
		\label{eqn:BC_at_NE}
		-\boldsymbol{ n}\cdot d_X\boldsymbol{\nabla}X_n(\boldsymbol{r},t) + \mathcal{J}_X = 0,\;\; \boldsymbol{r}\in\mathcal{S},
	\end{equation} where $\boldsymbol{n}$ is the outward unit normal to the NE surface $\mathcal{S}$.
	
	We consider the dynamics of the Ran cycle as shown in Fig.~\ref{fig:figure_1}, where $A$ denotes RanGDP, $B$ is RanGTP, and $I$ is RanGTP--NTR complex. The dynamics of the nuclear RanGDP is given by the reaction--diffusion equation:
	\begin{equation}
		\label{eqn:dyn-A}
		{\partial_t A_n}= d_{A}\nabla^2\hspace{-1pt} A_n - \alpha(\boldsymbol{r},t) A_n,\;\;\boldsymbol{r}\in\mathcal{V}_n,
	\end{equation} where $\alpha$ is the nucleotide exchange rate due to RanGEF, converting RanGDP to RanGTP~\footnote{ RanGDP is transported into the nucleus by binding to Nuclear Transport Factor 2 (NTF2). Here we do not explicitly consider the interactions with NTF2. We assume that the rate of nucleotide exchange for Ran is not limited by NTF2, assuming the RanGDP molecule dissociates rapidly from that complex once it enters the nucleus; though its dissociation mechanism still remains elusive.}. Similarly, the dynamics of nuclear RanGTP is described by
	\begin{equation}
		\label{eqn:dyn-B}
		{\partial_t B_n} = d_{B}\nabla^2\hspace{-1pt} B_n + \alpha(\boldsymbol{r},t) A_n - \beta(\boldsymbol{r},t)B_n,\;\;\boldsymbol{r}\in\mathcal{V}_n,
	\end{equation} where $\beta$ is the dissociation rate of cargo molecules from the cargo--NTR complex by binding of RanGTP. We assume $\beta(\boldsymbol{r},t) = b_0 G_n(\boldsymbol{r},t)$, where $b_0$ is a constant and $G_n$ is the nuclear cargo--NTR concentration given by
	\begin{equation}
		\label{eqn:dyn-G}
		{\partial_t G_n} = d_{G}\nabla^2\hspace{-1pt} G_n - b_0 G_n B_n,\;\;\boldsymbol{r}\in\mathcal{V}_n.
	\end{equation} Also, the cargo dissociation leads to the formation of a nuclear RanGTP--NTR complex ($I_n$), which is given by 
	\begin{equation}
		\label{eqn:dyn-I}
		{\partial_t I_n} = d_{I}\nabla^2\hspace{-1pt} I_n + \beta(\boldsymbol{r},t)B_n, \;\;\boldsymbol{r}\in\mathcal{V}_n.
	\end{equation} The nuclear concentration $C_n$ of free cargo  follows
	\begin{equation}
		\label{eqn:dyn-C}
		{\partial_t C_n} = d_{C}\nabla^2\hspace{-1pt} C_n + \beta(\boldsymbol{r},t)  B_n - \chi(\boldsymbol{r},t) C_n,\;\,\boldsymbol{r}\in\mathcal{V}_n,
	\end{equation} where $\chi$ is the nuclear depletion of cargo, e.g.~the transcription factor binding to certain loci on the chromatin. 

	\begin{figure*}[t!]\includegraphics[width=\textwidth]{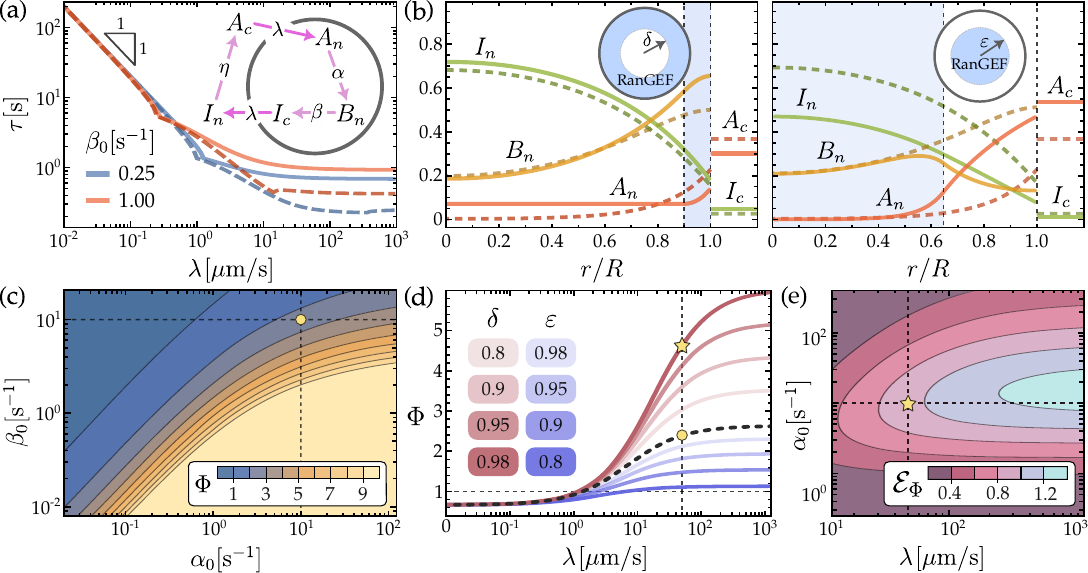}
	\caption{\label{fig:figure_2} (a) Relaxation time $\tau$ of the Ran cycle (see diagram), when RanGEF is uniform with $\alpha_0=5\,\mathrm{s}^{-1}$. Herein, we set the same diffusivity $d=10\,\mu\mathrm{m}^2/\mathrm{s}$ for all species, $R=10\,\mu\mathrm{m}$, $\eta=10\,\mathrm{s}^{-1}$, and $\nu={2}/{3}$. Dashed curves are the associated infinite diffusivity limits (same colors). (b) Steady-state concentrations of the Ran-cycle components (as fraction of initial $A_0$ concentration). Dashed curves correspond to a uniform RanGEF. In the left plot, solid curves correspond to RanGEF in a shell with inner radius $R\hspace{1pt}\delta$ (blue region). RanGEF's net number is the same as in the uniform case, $\alpha_\mathrm{shell} = \alpha_0/(1-\delta^3)$.  In the right plot, RanGEF is in a ball of radius $R\hspace{1pt}\varepsilon$ (blue region), with $\varepsilon$ such that $\alpha_\mathrm{ball} = \alpha_0/\varepsilon^3 = \alpha_\mathrm{shell} $. Here, $\delta=0.9$, $\beta_0 = 1\,\mathrm{s}^{-1}\!$, and $\lambda=10\,\mu\mathrm{m}/\mathrm{s}$. (c)~Steady-state $\Phi$ when RanGEF is uniform, with $\lambda=50\,\mu\mathrm{m}/\mathrm{s}$. (d)~Steady-state $\Phi$ for the uniform (dashed), shell (purple), and ball (blue) cases, with $\alpha_0=10\,\mathrm{s}^{-1}\!$ and $\beta_0=1\,\mathrm{s}^{-1}\!$. (e) Relative change $\mathcal{E}_\Phi$ of the steady-state $\Phi$ in the shell case compared to $\Phi$ in the uniform case, with $\delta=0.98$. Yellow point and star in (c--e) highlight the same parameter points.}
	\end{figure*}
	
	The five reaction--diffusion equations (\ref{eqn:dyn-A}--\ref{eqn:dyn-C}) are augmented with boundary conditions at the NE, which are given by Eq.~(\ref{eqn:BC_at_NE}) with $\mathcal{J}_A$, $\mathcal{J}_G$, and $\mathcal{J}_I$ being nonzero, since their constituents associate with NTRs, while $\mathcal{J}_B=\mathcal{J}_C=0$, as $B$ and $C$ cannot translocate through NPCs. 
	
	In cytoplasm, the homogeneous RanGTP--NTR concentration is described by
	\begin{equation}
		\label{eqn:dyn-I-out}
		{\partial_t I_c} = - \eta I_c-{\mathcal{V}_c^{-1}}\!\textstyle\int_{\mathcal{S}}\hspace{1pt}{\mathcal{J}_I}\,\mathrm{d}S,
	\end{equation} where the last term is the exchange of RanGTP--NTR through the NPCs, with $\mathcal{V}_c$ as the volume of cytoplasm, whereas the first is the loss due to the GTP hydrolysis by RanGAP, converting RanGTP to RanGDP at a rate $\eta$. Similarly, the cytoplasmic RanGDP is governed by
	\begin{equation}
		\label{eqn:dyn-A-out}
		{\partial_t A_c} = \eta I_c -{\mathcal{V}_c^{-1}}\!\textstyle\int_{\mathcal{S}}\hspace{1pt}{\mathcal{J}_A}\,\mathrm{d}S,\\[5pt]
	\end{equation} while the cytoplasmic cargo--NTR complex follows
	\begin{equation}
		\label{eqn:dyn-G-out}
		{\partial_t G_c} = \zeta_0 N_c  C_c-{\mathcal{V}_c^{-1}}\!\textstyle\int_{\mathcal{S}}\hspace{1pt}{\mathcal{J}_{G}}\,\mathrm{d}S,
	\end{equation} where $N_c$ is the cytosolic concentration of free NTRs, and  $\zeta_0$ is the production rate of cargo--NTR complex from a pool of free cargo, with $\partial_tC_c = -\zeta_0 N_c C_c \;+ $ other sources or sink terms. NTRs are assumed to be abundant in the cell with $N_c$ as a constant. Eqs.~(2--9) can be solved by providing initial conditions for each concentration.
	\section{Ran cycle under static nuclear RanGEF profiles} 
	\begin{figure*}[t!]\includegraphics[width=\textwidth]{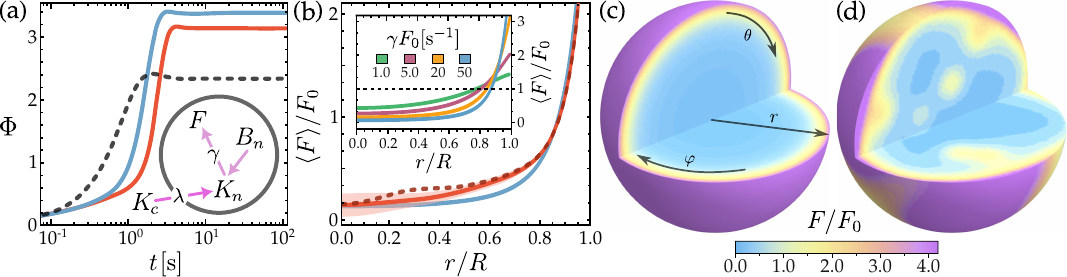}
	\caption{\label{fig:figure_3} (a) Relaxation of $\Phi$ when RanGEF concentration  is uniform (dashed), $\alpha = a_0 F_0 =3\,\mathrm{s}^{-1}$. Solid curves are the cases when RanGEF is transported in the nucleus as a complex $K$ (see diagram) and dissociated at a rate $\gamma F_0  = 50\,\mathrm{s}^{-1}$. Initial concentration of $F$ is taken to be: uniform (blue curve), and heterogeneous given by 10 random Gaussian spherical clusters of equal size. Red curve is $\Phi$ averaged over 100 such random initial data. (b) The angular average $\langle F\rangle$ of RanGEF at steady-state. Blue and red curves corresponds to the uniform and random initial cases as in (a). Inset shows the dependence on $\gamma$ with RanGEF initially uniform. The blue curve is also depicted in (c). The red shaded region is the standard deviation of the sample around the mean. The red dashed shows one such realization, which is also depicted in figure (d) and its initial data is shown in Appendix~C (inset plot of Fig.~10). 
	}
	\end{figure*}
	Since the Ran cycle drives and regulates the nucleocytoplasmic transport, we aim to understand its dynamics and steady-state behavior. We assume that cargo--NTR complexes are in abundance within the cell, so that the cargo dissociation rate $\beta=\beta_0$ is treated as a constant. To start, we assume that RanGEF is uniformly distributed with concentration $F_0$, where $\alpha= a_0 F_0 \equiv \alpha_0$ and $a_0$ is a constant. Under this approximation, the equations become linear and can be solved by a Laplace transform method {(Appendix~A.1)}, choosing an initial condition where Ran is only in the cytoplasm in its GDP-form. The solutions in  Laplace space can be used to determine the steady-state concentration profiles and the dominant relaxation time to steady-state {(Appendix~A.2)}, as shown in Fig.~\ref{fig:figure_2}(a) and (b), where, for simplicity, we choose the permeability $\lambda$ and diffusivity $d$ to be the same for all species. Since initially Ran molecules are not present in the nucleus and cytoplasmic concentrations are homogeneous, the solutions to the nuclear concentrations at steady state are found to be radially symmetric, decreasing from the NE into the nucleus with permeation length scales $\ell_A = \sqrt{d/\alpha_0}$ and $\ell_B = \sqrt{d/\beta_0}$. These lengths also control the abundance of molecules in the nucleus compared to cytoplasm, where the ratio $\Phi$ of nuclear-to-cytoplasmic Ran molecules is found in exact form  {(see Appendix~A.3)}. 
	In the physiological regime, $\Phi$ must be greater than unity, around $3$--$4$ \cite{Abu-Arish2009a}. This condition restricts the phase space in terms of local rates, see Fig.~\ref{fig:figure_2}(c),  requiring that $\beta_0$ be less than a threshold $\beta_0^{\star}$ at which $\Phi=1$, with typically $\beta_0^\star\simeq1\,\mathrm{s}^{-1}$. Similarly, this further constrains the permeability~$\lambda$, which must be larger than a threshold $\lambda^\star$, see Fig.~\ref{fig:figure_2}(d); typically $\lambda_\star\simeq1\,\mu\mathrm{m}/\mathrm{s}$. Note that if the rates $\alpha_0$ and $\beta_0$ were known, then knowledge of $\Phi$, which can be measured experimentally~\cite{Abu-Arish2009a},  allows the estimation of biophysical parameters such as permeability~$\lambda$.
	Another experimentally measurable quantity \cite{Timney2016,Timney2006,Gorlich2003,Kopito2007, Kopito2009,Riddick2007,Ribbeck2001} is the relaxation time to steady-state of the Ran cycle. In this simple theory, the relaxation time follows from the Laplace space solutions of the concentrations {(Appendix~A.4)}. Specifically, the complex pole $s_\star$ with the smallest negative real part in this solution gives us the dominant relaxation time $\tau$; shown in Fig.~\ref{fig:figure_2}(a). For small values of permeability $\lambda$, we find that $\tau\sim1/\lambda$, while at larger values of $\lambda$, the system exhibits underdamped oscillations (frequency  given by a nonzero imaginary part of $s_\star$), where oscillation onset occurs at higher $\lambda$ with increasing $\beta$. As shown in {Appendix~A.4}, a comparison between $\tau$ and its value in the limit of fast diffusion shows large deviations for $\lambda\gtrsim1\,\mu\mathrm{m}/\mathrm{s}$, and thus neglecting diffusion may lead to significant errors in the estimation of the relaxation times.\\
	\indent
	The nucleotide exchange rate depends on the local concentration of RanGEF bound to the chromatin~\cite{Tachibana2000}, which so far is assumed to be uniformly distributed. However, chromatin density is heterogeneous~\cite{Shivashankar2006}, including heterochromatin. Hence, we investigate next the effect of the RanGEF spatial heterogeneity \cite{Paschal2019}. For this we study two simple radial distributions of $\alpha(r)$: first, all RanGEF molecules are localized in a spherical shell; and, second, they are all confined to a spherical ball at the center of the nucleus. Assuming piece-wise constant profiles 
	of $\alpha$, the steady-state radial profiles of the concentration fields can be found analytically {(see Appendix~B.1 and B.2)}, and are shown in Fig.~\ref{fig:figure_2}(b) for the shell and ball cases. {The shell thickness is given by $R(1-\delta)$, whereas the radius of the spherical ball is $R\varepsilon$, as depicted in the inset diagrams of Fig.~\ref{fig:figure_2}(b).} In both scenarios, the same total number of RanGEF molecules is chosen as in the entirely uniform case with $\alpha =\alpha_0$. The associated steady-states of $\Phi$ can also be determined in exact form {(see Appendix~B.3)}, which are plotted in Fig.~\ref{fig:figure_2}(d). This shows that distributing RanGEF in a shell near the NE significantly increases the nuclear localization of molecules, when compared to the uniform case; see Fig.~\ref{fig:figure_2}(e). The increase in $\Phi$ increases with diminishing the shell thickness. Conversely, localizing RanGEF away from the NE (the ball case) has the opposite effect. Thus, a spatial profile for RanGEF can control  the nuclear transport.
	
	\section{Transport of RanGEF as a nuclear cargo} 
	
	So far we have assumed static RanGEF distributions, however, RanGEF molecules are also NLS-containing cargoes and their nuclear transport is mediated by the Ran cycle~\cite{Sankhala2017}. Next, we show that within our model this results in a positive feedback that localizes RanGEF to the nuclear periphery.
	To begin, we complement the previous equations of the Ran cycle with  transport of an additional cargo---receptor complex $K$ that carries RanGEF ($F$); see inset diagram in Fig.~\ref{fig:figure_3}(a). 	By using Eqs.~(\ref{eqn:dyn-G}) and (\ref{eqn:dyn-G-out}), their cytoplasmic and nuclear concentration are described by  
	\begin{equation}
		\partial_t K_c = -{\mathcal{V}_c^{-1}}\!\int_{\mathcal{S}}\mathcal{J}_K\,\mathrm{d}S,
	\end{equation}
	\begin{equation}
		\partial_t K_n  = d_K \nabla^2\hspace{-1pt} K_n - \gamma B_n K_n,
	\end{equation} respectively, satisfying a flux boundary condition as in Eq.~(\ref{eqn:BC_at_NE}). The last (nonlinear) term accounts for dissociation of RanGEF cargo at the rate $\gamma B_n$. We assume that once free the RanGEF cargo binds rapidly to the chromatin, and its bound concentration is given by 
	\begin{equation}
		\partial_t F  = \gamma B_n K_n.
	\end{equation} To complete the model we must set the rates to
	\begin{equation}
		\alpha = a_0 F\quad\text{and}\quad\beta = \beta_0 + \gamma K_n.
	\end{equation} 
	
	First, we consider initial distributions $F$ for bound RanGEF which comprise only 10\% of the total RanGEF in the cell. The remaining 90\% is in the cytoplasm which sets the initial data for $K_c$. As before, all of the Ran molecules are initially in the cytoplasm and in their GDP-form. We solve numerically for the spatial-temporal evolution of the concentration fields \cite{Burns2020}{, see Appendix~C}. 
	Fig.~\ref{fig:figure_3}(a) shows the evolution of $\Phi$ for (i) an initially uniform distribution for $F$, and (ii) a distribution $F$ consisting of randomly placed Gaussian spherical clusters of equal standard deviation. 
	The long--time states of $\Phi$ for both (i) and (ii) show an significant increase over the static uniform RanGEF case, as previously computed in Fig.~\ref{fig:figure_2}(d). Moreover, the long--time concentration profile of chromatin-bound RanGEF shows a sharply varying spatial profile within the nucleus, with a significant accumulation at the NE.	To compare the cases (i) and (ii), we compute the angular average of RanGEF,
	\begin{equation}
		\langle F\rangle = \frac{1}{4\pi}\int^{2\pi}_0\mathrm{d}\varphi\int^{\pi}_{0} F(r,\theta,\varphi)\sin\theta\mathrm{d}\theta,
	\end{equation}
	which are shown in Fig.~\ref{fig:figure_3}(b). For case (i), we find at long times a radial profile $\langle F\rangle$ that monotonically decays away from NE with a permeation length $\sim\!\sqrt{d/(\gamma F_0)}$, which rapidly decreases by increasing the rate $\gamma$; see Fig.~\ref{fig:figure_3}(b) and (c). For case (ii), the long-time $F$ shows a heterogeneous profile, as shown in the example of Fig.~\ref{fig:figure_3}(d), where $\langle F\rangle$ also sharply decays away from the NE. By averaging over many such radial profiles $\langle F\rangle$ at long--times, each derived with different initial random data {(see Appendix~C)}, we determine their mean and standard deviation as shown in Fig.~\ref{fig:figure_3}(b). This reveals that bound RanGEF displays a significant and sharp accumulation at the NE, despite the initial random distribution. 
	
	{Similarly, we can compare the spatiotemporal dynamics of the other Ran components under the initial conditions (i) and (ii). The angular average concentrations of nuclear RanGDP, RanGTP, and RanGTP--NTR are shown in Appendix~C (see Fig.~\ref{fig:grid-plots-abi}), which shows heterogeneous radial profiles similar to cases in Fig.~\ref{fig:figure_2}~(b) where RanGEF is static and spatially nonhomogeneous.}
	
	\vspace{-0.2cm}
	\section{Discussion and Summary} 
	
	We developed a model to explore nucleocytoplasmic transport, emphasizing the crucial role of the spatial distribution and dynamics of the Ran cycle. Our findings demonstrate that the spatial RanGEF distribution directly influences the nucleocytoplasmic transport, in particular, the localization of RanGEF at the NE increases the Ran content in the nucleus. We find that the sharp accumulation of RanGEF at the NE emerges inherently from the transport dynamics of RanGEF molecules as an NLS-containing cargo.\\
	\indent		
	{Our model predicts that reducing RanGEF activity, represented by a decrease in the parameter $\alpha$, lowers the nuclear-to-cytoplasmic Ran ratio, $\Phi$, see Fig.~\ref{fig:figure_2}(c). This reduction can be achieved by inhibiting RanGEF, thus diminishing its nuclear concentration $F$, a result corroborated by several experimental studies \cite{Tachibana1994, Hood2007, Paschal2011, Paschal2014, Paschal2015}. Additionally, our predictions suggest that reducing the nuclear pore permeability $\lambda$ associated with Ran can further decrease $\Phi$, see Fig.~\ref{fig:figure_2}(d).\\
	\indent		
	A key prediction of our model is that the spatial distribution of RanGEF within the nucleus, even if the net number of RanGEF molecules remains unchanged, significantly influences the $\Phi$ ratio. Specifically, redistributing RanGEF closer to the nuclear periphery markedly increases $\Phi$, while homogenizing RanGEF from its spatially nonuniform state results in a decrease in $\Phi$, see Fig.~\ref{fig:figure_2}(e). This is consistent with experimental perturbations that disrupt the chromatin organization. In particular, RanGEF has a higher affinity for heterochromatin \cite{Casolari2004} which is located primarily at the nuclear boundary. A reduction in $\Phi$ has been experimentally observed by decreasing either the levels of heterochromatin markers or HP1 \cite{Paschal2011, Paschal2014, Paschal2015, Paschal2019}. Furthermore, disrupting nuclear lamin, as seen in progerin patient cell lines where both lamin disruption and heterochromatin alterations are evident, also reduces $\Phi$ \cite{Paschal2011, Paschal2014, Paschal2015, Paschal2019}, being an example of a human disease that displays a disruption of the Ran gradient.}\\
	\indent	
	Overall, our results suggest that the spatial modulation of the RanGEF substrate may contribute to the regulation of the RanGTP spatial gradient inside cells. With RanGEF having a high affinity for heterochromatin, changes in its organization near the NE could affect the spatial dynamics of RanGEF, and thus in turn the nucleocytoplasmic transport. Understanding how changes in heterochromatin influence RanGEF localization may offer insights into cell's progression through the cell cycle as well as diseases, where nucleus-wide chromatin organization is disrupted. Our results open new avenues for further theoretical and experimental inquiries into the role of spatial dynamics of molecules in the nucleocytoplasmic transport. 
	
	\begin{acknowledgments}
		{The authors gratefully acknowledge funding from National Science Foundation Grants No.\ CMMI-1762506 and No.\ DMS-2153432 (A.\ Z.\ and M.\ J.\ S.), No.\ CAREER PHY-1554880 and No.\ PHY-2210541 (A.\ Z.). The authors also thank D.\ Saintillan, J.\ Alsous, A.\ Lamson, S.\ Weady, and B.\ Chakrabarti for stimulating discussions and helpful feedback.}
	\end{acknowledgments}
	
	\section*{Appendix}\appendix
	
	\section{Homogeneous RanGEF distribution} 
	
	The cargo dissociation rate $\beta=\beta_0$ is taken as a constant; assuming that NTR--cargo complexes are found in excess throughout the cell. Also, RanGEF is homogeneously distributed within the nucleus: $\alpha=\alpha_0 = a_0 F_0$, where $F_0$ is the concentration of RanGEF, and $a_0$ is a constant. Under such assumptions, the system of equations becomes linear and thus their solutions are analytically tractable.	In the following, we choose for simplicity that nuclear diffusivity of all Ran species is the same, denoted by $d$. The permeability of RanGDP and NTR--RanGTP is chosen to be the same, denoted by $\lambda$.  
	
	\subsection{Radial solutions in Laplace space} 
	
	Provided that the initial conditions for the nuclear concentrations are uniform, their steady-state solutions will remain spherically symmetric as the cytoplasmic concentrations are treated as spatially homogeneous. The nuclear concentration of RanGDP can be described by
	\begin{equation}
		\label{eqn:dyn-A-linear}
		T \frac{\partial A_n}{\partial t} = \frac{1}{r^2}\frac{\partial}{\partial r}\!\left[r^2\frac{\partial A_n}{\partial r}\right]\!- k_{\alpha}  A_n(r,t),
	\end{equation} where $T = R^2/d$ is the diffusion time, the dimensionless nucleotide exchange rate $k_{\alpha} = R^2\alpha_0/d$, and $r\in[0,1]$ is the rescaled radial distance from the center of the cell nucleus ($r=0$). Moreover,  Eq.~(\ref{eqn:dyn-A-linear}) satisfies the following flux boundary condition at the NE ($r=1$):
	\begin{equation}
		\label{eqn:flux-A-linear}
		\frac{\partial A_n}{\partial r}(r\!=\!1,t) = \Lambda \left[A_{c}(t) - A_n(1,t)\right]\!,
	\end{equation} where $\Lambda =  \lambda R/d$. Initially, we consider that RanGDP is not present in the nucleus, that is, $A_n(r,t\!=\!0) = 0$.  The cytoplasmic dynamics of $A_c$ is given by
	\begin{equation}
		\label{eqn:dyn-Ac-linear}	
		T\frac{\partial A_{c}}{\partial t} = k_\eta I_c-3\nu{\Lambda}\left[A_{c}(t) - A_n(1,t)\right]\!,
	\end{equation} where the  dimensionless recycling rate $k_\eta = \eta R^2/d$ and $\nu = \mathcal{V}_n/\mathcal{V}_c$. We set $A_c(t\!=\!0)=A_0$, with $A_0$ as the initial concentration of Ran molecules in the cytoplasm.
	
	Similarly, the nuclear RanGTP is governed by
	\begin{equation}
		\label{eqn:dyn-B-linear}
		T \frac{\partial B_n}{\partial t} = \frac{1}{r^2}\frac{\partial}{\partial r}\!\left[r^2\frac{\partial B_n}{\partial r}\right]\!+ k_{\alpha} A_n-k_{\beta} B_n,
	\end{equation} where $k_{\beta} = R^2\beta_0/d$ is the dimensionless cargo dissociation rate. The RanGTP is initially absent in the nuclues, that is, $B(r,t\!=\!0)=0$. Furthermore, $B_n$ satisfies a no-flux boundary condition at the NE:
	\begin{equation}
		\label{eqn:flux-B-linear}
		\frac{\partial B_n}{\partial r}(r\!=\!1,t) = 0.
	\end{equation} 
	
	The concentration of NTR--RanGTP complexes in the nucleus is given by
	\begin{equation}
		\label{eqn:dyn-I-linear}
		T \frac{\partial I_n}{\partial t} = \frac{1}{r^2}\frac{\partial}{\partial r}\!\left[r^2\frac{\partial I_n}{\partial r}\right]\!+k_{\beta}  B_n,
	\end{equation} which satisfies the following boundary condition at NE:
	\begin{equation}
		\label{eqn:flux-I-linear}
		\frac{\partial I_n}{\partial r}(r\!=\!1,t) = \Lambda \left[I_{c}(t) - I_n(1,t)\right]\!,
	\end{equation} and the cytoplasmic concentration $I_c$ is governed by
	\begin{equation}
		\label{eqn:dyn-Ic-linear}	
		T\frac{\partial I_{c}}{\partial t} = -k_\eta I_c-3\nu{\Lambda}\left[I_{c}(t) - I_n(1,t)\right]\!.
	\end{equation} Here, we choose $I_n(r,t\!=\!0) = 0$ and $I_c(t\!=\!0)=0$; thus, all of the Ran molecules are initially in the cytoplasm in the RanGDP-form (at concentration $A_0$).
	
	The set of equations (\ref{eqn:dyn-A-linear}--\ref{eqn:dyn-Ic-linear}) can be solved by a Laplace transform method, where we define that
	\begin{equation}
		\label{eqn:Laplace-transform-def}
		\hat{X}(s)=\hat{\mathcal{L}_s}[{X}] = \frac{1}{T}\int_{0}^{\infty}\!e^{-st/T}\,X(t)\,\mathrm{d}t, 
	\end{equation} where $X$ denotes any of the concentrations ${A}_n$, ${A}_c$, $B_n$, ${I}_n$, and ${I}_c$. By employing Eq.~(\ref{eqn:Laplace-transform-def}), the nuclear RanGDP equation (\ref{eqn:dyn-A-linear}) can be written as 
	\begin{equation}
		\label{eqn:dyn-A-linear-LT}
		\frac{1}{r^2}\frac{\partial}{\partial r}\!\bigg[r^2\frac{\partial \hat{A}_n}{\partial r}\bigg]\!- (s+k_{\alpha})\hspace{1pt}\hat{A}_n(r,s) = 0,
	\end{equation} and the boundary condition in Eq.~(\ref{eqn:flux-A-linear}) becomes
	\begin{equation}
		\label{eqn:flux-A-linear-LT}
		\frac{\partial \hat{A}_n}{\partial r}(r\!=\!1,s) = \Lambda\,[\hat{A}_{c}(t) - \hat{A}_n(1,s)].
	\end{equation} The cytoplasmic RanGDP equation  Eq.~(\ref{eqn:dyn-Ac-linear}) becomes
	\begin{equation}	
		s\hat{A}_{c}(s)-A_0 = k_\eta \hat{I}_c(s)-3\nu{\Lambda}[\hat{A}_{c}(s) - \hat{A}_n(1,s)], 
	\end{equation} which gives
	\begin{equation}
		\label{eqn:hat-A-c}
		\hat{A}_{c}(s) = \frac{A_0 + k_\eta \hat{I}_c(s)+3\nu{\Lambda}\hat{A}_n(1,s)}{s+3\nu{\Lambda}}.
	\end{equation}
	
	The solution of Eq.~(\ref{eqn:dyn-A-linear-LT}) can be found to be
	\begin{equation}
		\label{eqn:hat-A-n}
		\hat{A}_n(r,s) = \frac{\mathcal{C}_{\alpha}(s)\sinh(r \mathcal{S}_{\alpha})}{r},
	\end{equation} where $\mathcal{S}_{\alpha} = \sqrt{s+k_{\alpha}}$ and $\mathcal{C}_{\alpha}(s)$ is a function to be prescribed by the boundary condition in Eq.~(\ref{eqn:flux-A-linear-LT}); namely,
	\begin{equation}
		\label{eqn:C_alpha}
		\mathcal{C}_{\alpha}(s) = \frac{\hat{A}_c(s)\hspace{1pt}\Lambda}{ \mathcal{S}_{\alpha}\cosh (\mathcal{S}_{\alpha})-(1-\Lambda)\sinh(\mathcal{S}_{\alpha})}.
	\end{equation} 
	
	By Laplace transforming Eq.~(\ref{eqn:dyn-B-linear}), the equation of  nuclear RanGTP can be written as 
	\begin{equation}
		\label{eqn:dyn-B-linear-LP}
		\frac{1}{r^2}\frac{\partial}{\partial r}\!\bigg[r^2\frac{\partial \hat{B}_n}{\partial r}\bigg]\!+ k_{\alpha}\hat{A}_n-(s+k_{\beta}) \hat{B}_n = 0,
	\end{equation} and its no-flux boundary condition now reads
	\begin{equation}
		\label{eqn:flux-B-linear-LT}
		\frac{\partial }{\partial r}\hat{B}_n(r\!=\!1,t) = 0.
	\end{equation} 
	
	Using Eq.~(\ref{eqn:hat-A-c}), the solution of Eq.~(\ref{eqn:dyn-B-linear-LP}) is given by
	\begin{equation}
		\label{eqn:hat-B-n}
		\hat{B}_n(r,s) = \frac{\mathcal{C}_{\beta}(s)\sinh(r \mathcal{S}_{\beta})}{r}+\frac{k_{\alpha}\hspace{1pt}\mathcal{C}_{\alpha}(s)\sinh(r \mathcal{S}_{\alpha})}{r\hspace{1pt}(k_{\beta}-k_{\alpha})},
	\end{equation} where we define $\mathcal{S}_{\beta} = \sqrt{s+k_{\beta}}$, and $\mathcal{C}_{\beta}$ is a function to be found from the boundary condition in Eq.~(\ref{eqn:flux-B-linear-LT}):
	\begin{equation}
		\label{eqn:C_beta}
		\mathcal{C}_{\beta}(s) = \frac{\displaystyle \hat{A}_c(s)\left[\mathcal{S}_{\beta} \cosh \left(\mathcal{S}_{\beta}\right)-\sinh \left(\mathcal{S}_{\beta}\right)\right]^{-1} }{\displaystyle\left(1-\frac{k_{\beta}}{k_{\alpha}}\right)\!\! \left[\frac{1}{\Lambda}-\frac{1}{1-\mathcal{S}_{\alpha} \coth \left(\mathcal{S}_{\alpha}\right)}\right]}.
	\end{equation}
	
	By Laplace transforming the nuclear NTR--RanGTP equation (\ref{eqn:dyn-I-linear}), we obtain that
	\begin{equation}
		\label{eqn:dyn-I-linear-LP}
		\frac{1}{r^2}\frac{\partial}{\partial r}\!\bigg[r^2\frac{\partial \hat{I}_n}{\partial r}\bigg]\!+ k_{\beta}\hat{B}_n-s\,\hat{I}_n = 0,
	\end{equation} while the boundary condition in Eq.~(\ref{eqn:flux-A-linear}) transforms to
	\begin{equation}
		\label{eqn:flux-I-linear-LT}
		\frac{\partial \hat{I}_n}{\partial r}(r\!=\!1,s) = \Lambda\,[\hat{I}_{c}(t) - \hat{I}_n(1,s)].
	\end{equation} The solution to Eq.~(\ref{eqn:dyn-I-linear-LP}) can be found as follows:
	\begin{align}
		\label{eqn:hat-I-n}
		\hat{I}_n(r,s) &= \frac{\mathcal{C}_{\alpha}(s) k_{\beta} \sinh \left(r \mathcal{S}_{\alpha}\right)}{r \left(k_{\alpha}-k_{\beta}\right)}-\frac{\mathcal{C}_{\beta}(s) \sinh \left(r \mathcal{S}_{\beta}\right)}{r}\notag\\[2pt]
		&\quad+\frac{\mathcal{C}_0(s) \sinh
			\left(r \sqrt{s}\right)}{r},
	\end{align} where $\mathcal{C}_0$ is a function to be determined from the boundary condition in Eq.~(\ref{eqn:flux-I-linear-LT}); namely, we find that
	\begin{equation}
		\label{eqn:C_0}
		\mathcal{C}_0(s) = \frac{\displaystyle \hat{I}_c(s)\hspace{1pt}\Lambda + \hat{A}_c(s)\hspace{1pt}\Lambda  \!\left[1-\frac{1-{\mathcal{X}_{\beta}}/{\mathcal{X}_{\alpha}}}{1-{k_{\beta}}/{k_{\alpha}}}\right]}{(\Lambda -1) \sinh
			\left(\sqrt{s}\right)+\sqrt{s} \cosh \left(\sqrt{s}\right)},
	\end{equation} where we define
	\begin{equation}
		\mathcal{X}_j=\frac{1}{\Lambda}-\frac{1}{1-\mathcal{S}_j \coth \left(\mathcal{S}_j\right)},\quad j\in\{\alpha,\beta\}.
	\end{equation}
	
	From Eq.~(\ref{eqn:dyn-Ic-linear}), the cytoplasmic concentration of the NTR--RanGTP complex in Laplace space is found to be 
	\begin{equation}
		\label{eqn:hat-I-c}
		\hat{I}_{c}(s) =\frac{3\nu{\Lambda}}{s+k_\eta+3\nu{\Lambda}}\,\hat{I}_n(1,s).
	\end{equation} This equation allows us to obtain $\hat{A}_c(s)$ from Eq.~(\ref{eqn:hat-A-c}), which in turn provides us with exact expressions in the Laplace space for the concentrations $\hat{A}_n(r,s)$, $\hat{B}_n(r,s)$, and $\hat{I}_n(r,s)$ in terms of the initial concentration $A_0$ and the dimensionless parameters $\nu$, $k_{\alpha}$, $k_{\beta}$, $k_\eta$, and $\Lambda$.
	
	\subsection{Steady-state concentration profiles} 
	
	The steady-state concentration profiles can be derived by final value theorem for the Laplace transform, which tells us that if the steady-state value exists, then $$\lim_{t\rightarrow\infty} f(t) = \lim_{s\rightarrow0}s\hat{F}(s)$$ where $\hat{F}(s)$ is the Laplace transform of the function $f(t)$, provided that all poles of $s\hat{F}(s)$ are strictly stable or lie in the open left half-plane, $\mathrm{Re}(s)<0$.
	
	We start by assuming that the steady-state value of the cytoplasmic concentration $A_c(\infty)$ exists and we denote it hereinafter by $\bar{A}_c$. Therefore the steady-state nuclear concentration of RanGDP is given by
	\begin{equation}
		\label{eqn:bar-A-n}
		\bar{A}_n(r) = \frac{\bar{\mathcal{C}}_{\alpha}\sinh(r \sqrt{k_\alpha})}{r},
	\end{equation} where the constant $\bar{\mathcal{C}}_{\alpha}$ is found to be 
	\begin{equation}
		\bar{\mathcal{C}}_{\alpha} = \lim_{s\rightarrow0}s\,\mathcal{C}_{\alpha}(s) = \frac{\bar{A}_c\hspace{1pt}\Lambda\,\mathrm{csch}(\sqrt{k_\alpha})}{ \sqrt{k_\alpha}\coth (\sqrt{k_\alpha})-1+\Lambda},
	\end{equation}  through Eqs.~(\ref{eqn:hat-A-n}) and (\ref{eqn:C_alpha}), where $\bar{A}_c = \lim\limits_{s\rightarrow0}s\hat{A}_c(s)$. 

	Similarly, we find the steady-state of RanGTP concentration in the nucleus, $\bar{B}_n(r)$, from Eq.~(\ref{eqn:hat-B-n}), that is
	\begin{equation}
		\label{eqn:bar-B-n}
		\bar{B}_n(r) = \frac{\bar{\mathcal{C}}_{\beta}\sinh(r \sqrt{k_\beta})}{r}+\frac{k_{\alpha}\hspace{1pt}\bar{\mathcal{C}}_{\alpha}\sinh(r \sqrt{k_\alpha})}{r\hspace{1pt}(k_{\beta}-k_{\alpha})},
	\end{equation} where $\bar{\mathcal{C}}_{\beta}$ is readily found from Eq.~(\ref{eqn:C_beta}) as follows
	\begin{equation}
		\bar{\mathcal{C}}_{\beta} = \frac{\displaystyle \bar{A}_c\!\left[\sqrt{k_\beta} \cosh (\sqrt{k_\beta})-\sinh (\sqrt{k_\beta})\right]^{-1} }{\displaystyle\left(1-\frac{k_{\beta}}{k_{\alpha}}\right)\!\! \left[\frac{1}{\Lambda}-\frac{1}{1-\sqrt{k_\alpha} \coth (\sqrt{k_\alpha})}\right]}.
	\end{equation} 
	
	Also, the steady-state concentration profile of NTR--RanGTP in the nucleus, $\bar{I}_n(r)$, can be found in a similar way. Using Eqs.~(\ref{eqn:hat-I-n}) and (\ref{eqn:C_0}), we have that
	\begin{equation}
		\label{eqn:bar-I-n}
		\bar{I}_n(r) = \bar{\mathcal{C}}_0 + \frac{\bar{\mathcal{C}}_{\alpha} k_{\beta} \sinh (r \sqrt{k_\alpha})}{r \left(k_{\alpha}-k_{\beta}\right)}-\frac{\bar{\mathcal{C}}_{\beta} \sinh (r \sqrt{k_\beta})}{r},
	\end{equation} where the constant $\bar{\mathcal{C}}_0$ is given by
	\begin{equation}
		\label{eqn:bar-C_0}
		\bar{\mathcal{C}}_0 = \bar{I}_c+ \bar{A}_c \!\left[1-\frac{1-{\bar{\mathcal{X}}_{\beta}}/{\bar{\mathcal{X}}_{\alpha}}}{1-{k_{\beta}}/{k_{\alpha}}}\right]\!,
	\end{equation} with $\bar{I}_c = \lim\limits_{s\rightarrow0}s\hat{I}_c(s)$ and the coefficients 
	\begin{equation}
		\bar{\mathcal{X}}_j=\frac{1}{\Lambda}-\frac{1}{1-\sqrt{k_j} \coth (\sqrt{k_j})},\quad j\in\{\alpha,\beta\}.
	\end{equation} To derive the result in Eq.~(\ref{eqn:bar-C_0}), we use the limits
	\begin{align}
		&\lim\limits_{s\rightarrow0}\frac{\sinh(r\sqrt{s})}{\sqrt{s}} = r, \quad\text{and}\\
		&\lim\limits_{s\rightarrow0}\frac{\sqrt{s}}{(\Lambda -1) \sinh
			\left(\sqrt{s}\right)+\sqrt{s} \cosh \left(\sqrt{s}\right)} = \frac{1}{\Lambda}.
	\end{align}
	
	From Eq.~(\ref{eqn:hat-I-c}), we obtain that
	\begin{equation}
		\label{eqn:bar-I-c}
		\bar{I}_{c} =\frac{3\nu{\Lambda}}{k_\eta+3\nu{\Lambda}}\,\bar{I}_n(1) = \frac{3\nu \bar{A}_c}{k_\eta\bar{\mathcal{X}}_\alpha}.
	\end{equation} where the last equality follows from Eq.~(\ref{eqn:bar-I-n}). Using final value theorem, Eq.~(\ref{eqn:hat-A-c}) becomes
	\begin{equation}
		\label{eqn:bar-A-c}
		\bar{A}_{c} = \frac{k_\eta \bar{I}_c}{3\nu{\Lambda}}+\bar{A}_n(1).
	\end{equation} This also gives us that
	\begin{equation}
		\bar{A}_n(1) = \bar{A}_c\frac{{\Lambda\bar{\mathcal{X}}_\alpha}\!-\!1}{{\Lambda\bar{\mathcal{X}}_\alpha}} = \frac{\Lambda\bar{A}_c}{\sqrt{k_\alpha}\coth(\sqrt{k_\alpha})\hspace{-1pt}+\hspace{-1pt}\Lambda\hspace{-1pt}-\hspace{-1pt}1},
	\end{equation} which is consistent with Eq.~(\ref{eqn:bar-A-n}) at $r=1$. Moreover, the dimensionless flux of RanGDP at steady-state is 
	\begin{equation}
		\bar{\mathcal{J}}_A \equiv \Lambda\!\left[\bar{A}_{c} -\bar{A}_n(1)\right] = \frac{k_\eta \bar{I}_c}{3\nu}>0,
	\end{equation} which means that RanGDP diffuses into nuclues, while the dimensionless flux of NTR--RanGTP is given by
	\begin{equation}
		\bar{\mathcal{J}}_I \equiv \Lambda\!\left[\bar{I}_{c} -\bar{I}_n(1)\right] = -\frac{k_\eta \bar{I}_c}{3\nu}<0,
	\end{equation} which means that NTR--RanGTP diffuses out of the nucleus. Moreover, this shows that the fluxes must balance identically at steady-state:
	\begin{equation}
		\bar{\mathcal{J}}_A + \bar{\mathcal{J}}_I = 0. 
	\end{equation} 
	
	To obtain $\bar{A}_c$, we make use the number conservation of particles in the system, as described in the next section.
	
	\subsection{Nuclear-to-cytoplasmic ratio of molecules} 
	
	The number of cytoplasmic molecules in the form of RanGDP and NTR--RanGTP at time $t$ are given by 
	\begin{equation}
		\label{eqn:def-c}
		c_A(t) = \mathcal{V}_c\hspace{1pt}A_c(t),\quad\mathrm{and}\quad c_I(t) = \mathcal{V}_c\hspace{1pt} I_c(t),
	\end{equation} respectively, whereas the net number of nuclear RanGDP, $n_A(t)$, RanGTP, $n_B(t)$, and NTR--RanGTP, $n_I(t)$ are given by the integral of their corresponding concentration fields over the entire volume of the nucleus. Specifically, 
	\begin{align}
		\label{eqn:def-n-A}
		n_A(t) &= \int_{\mathcal{V}_n}\mathrm{d}V\,A_n(\boldsymbol{r},t),\\[2pt] 
		\label{eqn:def-n-B}
		n_B(t) &= \int_{\mathcal{V}_n}\mathrm{d}V\,B_n(\boldsymbol{r},t),\\[2pt]  
		\label{eqn:def-n-I}
		n_I(t) &= \int_{\mathcal{V}_n}\mathrm{d}V\,I_n(\boldsymbol{r},t).
	\end{align}

	By integrating over the nuclear volume the equations of $A_n$, $B_n$, and $I_n$, as shown in Eqs.~(2), (3), and (5), and using the divergence theorem and their respective boundary conditions given by Eq.~(1), we find that
	\begin{equation}
		\frac{\partial}{\partial t}\!\left(n_A+n_B+n_I\right) = \int_\mathcal{S}\mathrm{d}S\,\left(\mathcal{J}_A+\mathcal{J}_I\right)\!,
	\end{equation} where $\mathcal{S}$ is the NE. From the cytoplasmic concentration equations, see Eqs.~(7) and (8), we find
	\begin{equation}
		\frac{\partial}{\partial t}\!\left(c_A+c_I\right) = -\int_\mathcal{S}\mathrm{d}S\,\left(\mathcal{J}_A+\mathcal{J}_I\right)\!,
	\end{equation} which shows that the total number of molecules in the system is conserved, that is,
	\begin{equation}
		\label{eqn:conservation-0}
		\frac{\partial}{\partial t}\!\left(n_A+n_B+n_I + c_A+c_I\right) = 0.
	\end{equation} Since initially all of Ran molecules are assumed to be only in their GDP-form within the cytoplasm, with concentration $A_0$, we have that 
	\begin{equation}
		\label{eqn:conservation-1}
		n_A(t)+n_B(t)+n_I(t) + c_A(t)+c_I(t) = A_0\mathcal{V}_c,
	\end{equation} must be satisfied for all time $t$. This includes the steady-state, and therefore we require that
	\begin{equation}
		\label{eqn:conservation-2}
		\bar{n}_A + \bar{n}_B + \bar{n}_I  = \left(A_0 -\bar{A}_c-\bar{I}_c\right)\mathcal{V}_c
	\end{equation} where $\bar{n}_A$, $\bar{n}_B$ and $\bar{n}_I$ are the corresponding steady-state numbers. Using the steady-state concentration profile of nuclear RanGDP, as derived in Eq.~(\ref{eqn:bar-A-n}), we obtain
	\begin{equation}
		\label{eqn:bar-n-A}
		\bar{n}_A = 4\pi R^3\!\int^1_{0} \mathrm{d}r\,r^2\bar{A}_n(r) = \frac{3\hspace{1pt}\mathcal{V}_n\bar{A}_c}{k_\alpha\bar{\mathcal{X}}_\alpha},
	\end{equation}  where we use that $\mathcal{V}_n = 4\pi R^3/3$. Similarly, we can determine the number of nuclear RanGTP molecules at steady-state, which through Eq.~(\ref{eqn:bar-B-n}) yields
	\begin{equation}
		\label{eqn:bar-n-B}
		\bar{n}_B = 4\pi R^3\!\int^1_{0} \mathrm{d}r\,r^2\bar{B}_n(r) = \frac{3\hspace{1pt}\mathcal{V}_n\bar{A}_c}{k_\beta\bar{\mathcal{X}}_\alpha}.
	\end{equation} Note that number ratio of nuclear RanGTP to nuclear RanGDP at steady-state simply reduces to
	\begin{equation}
		\frac{\bar{n}_B}{\bar{n}_A} = \frac{k_\alpha}{k_\beta} = \frac{\alpha_0}{\beta_0}.
	\end{equation} 

	\begin{figure}[t!]\includegraphics[width=\columnwidth]{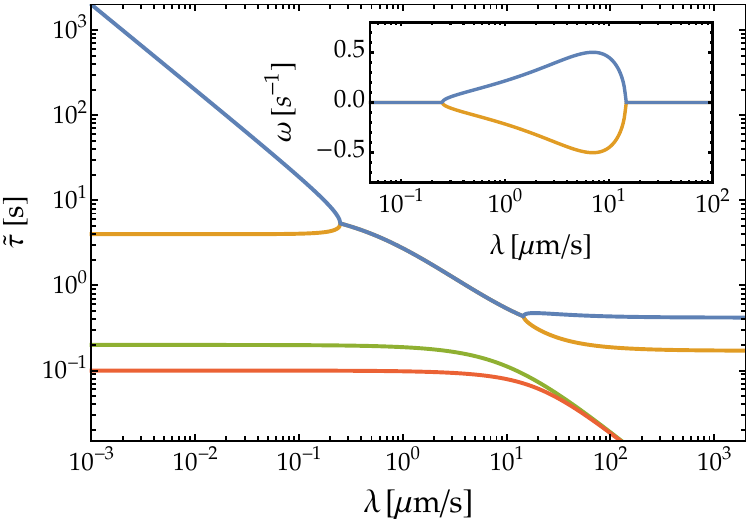}
	\caption{\label{fig:relaxation-times-fast-diffusion}  Relaxation times $\tilde{\tau}$ associated with each eigenvalue $\tilde{s}$ as function of permeability $\lambda$ for the fast diffusion case. The relaxation time is given by $\tilde{\tau} = -1/\mathrm{Re}(\tilde{s})$, while the frequency of underdamped oscillations is given by $\omega = \mathrm{Im}(\tilde{s})$, which is shown in the inset plot against $\lambda$. Here, we choose $\alpha_0=5\,\mathrm{s}^{-1}$, $\beta_0=0.25\,\mathrm{s}^{-1}$, $\eta=10\,\mathrm{s}^{-1}$, $R=10\,\mu\mathrm{m}$, and $\nu=2/3$.}
`	\end{figure}
	
	By means of Eq.~(\ref{eqn:bar-I-n}), the total number of nuclear NTR--RanGTP complexes is found to be
	\begin{equation}
		\label{eqn:bar-n-I}
		\bar{n}_I = \bar{\mathcal{C}}_0\mathcal{V}_n-\frac{3\hspace{1pt}\mathcal{V}_n(k_\alpha+k_\beta)\bar{A}_c}{k_\beta k_\alpha \bar{\mathcal{X}}_\alpha},
	\end{equation} where $\bar{\mathcal{C}}_0$ is given by Eq.~(\ref{eqn:bar-C_0}). Using Eqs.~(\ref{eqn:conservation-2}), (\ref{eqn:bar-n-A}), (\ref{eqn:bar-n-B}) and (\ref{eqn:bar-n-I}), we have that
	\begin{equation}
		\label{eqn:conservation-3}
		\nu\hspace{1pt}\bar{\mathcal{C}}_0 +\bar{A}_c + \bar{I}_c = A_0,
	\end{equation} which allows us to determine $\bar{A}_c$ in terms of $A_0$.
	
	Total nuclear-to-cytoplasmic ratio $\Phi$ of Ran molecules, which is defined by
	\begin{equation}
		\label{eqn:Phi-def}
		\Phi(t) = \frac{n_A(t)+n_B(t)+n_I(t)}{c_A(t)+c_I(t)},
	\end{equation} can be rewritten using Eq.~(\ref{eqn:conservation-1}) as follows:
	\begin{equation}
		\Phi(t) = \frac{A_0}{A_c(t)+I_c(t)}-1.
	\end{equation} Its steady-state, $\bar{\Phi}$, is further simplified using Eq.~(\ref{eqn:conservation-3}):
	\begin{equation}
		\label{eqn:Phi}
		\bar{\Phi} = \hspace{-1pt}\frac{\nu\hspace{1pt}\bar{\mathcal{C}}_0}{\bar{A}_c\hspace{-1pt}+\bar{I}_c}\hspace{-1pt}= \nu  \!\left[1\hspace{-1pt}-\hspace{-1pt}\frac{\bar{\mathcal{X}}_\alpha-\bar{\mathcal{X}}_\beta}{\left(1\hspace{-1pt}-\hspace{-1pt}{k_{\beta}}/{k_{\alpha}}\right)\!\left(\bar{\mathcal{X}}_\alpha\hspace{-1pt}+\hspace{-1pt}{3\nu }/{k_\eta}\right)}\right]
	\end{equation} where in the last equality we use Eq.~(\ref{eqn:C_0}) and (\ref{eqn:bar-I-c}). This also gives us a concise way to express in the steady-state value $\bar{A}_c$ in terms of $A_0$, namely
	\begin{equation}
		\bar{A}_c = \frac{A_0\,\bar{\mathcal{X}}_\alpha}{\left(1+\bar{\Phi}\right)\!\left(\bar{\mathcal{X}}_\alpha+{3\nu}/{k_\eta}\right)}.
	\end{equation} The steady-state value $\bar{I}_c$ becomes
	\begin{equation}
		\bar{I}_c = \frac{ A_0}{\left(1+\bar{\Phi}\right)\!\left[1+k_\eta\bar{\mathcal{X}}_\alpha/{(3\nu)}\right]},
	\end{equation} which shows that $\bar{I}_c$ vanishes in the limit $k_\eta\rightarrow\infty$, as expected; meaning the NTR-RanGTP complexes are instantaneously dissociated as they enter the cytoplasm.
	
	\subsection{Relaxation times to steady-state} 
	
	\begin{figure}[t!]\includegraphics[width=\columnwidth]{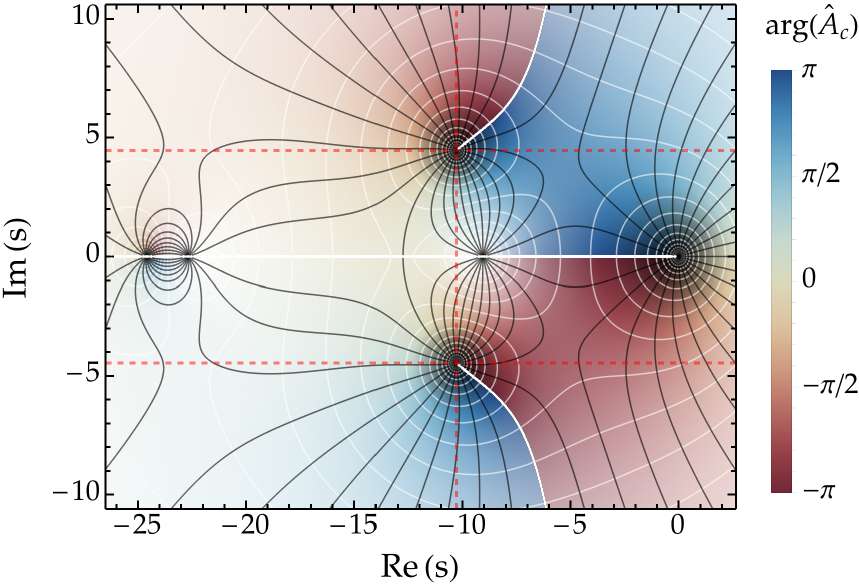}
	\caption{\label{fig:complex-plane}  Plot of $\hat{A}_c(s)$, with $s$ as a complex variable, where the color shows the $\arg(\hat{A}_c)$. Black and white meshes are contour lines of $\arg(\hat{A}_c)$ and $\log_{10}[\mathrm{abs}(\hat{A}_c)]$, respectively. Black shaded regions show the poles, while white shaded regions are the zeros of the function. The red dashed lines correspond to the poles with the smallest negative real part, which in this case also have a nonzero imaginary part. The white thick lines are branch cuts.  Herein, $\beta_0=0.25\,\mathrm{s}^{-1}\!$, $\alpha_0 =5\,\mathrm{s}^{-1}\!$, $\eta =10\,\mathrm{s}^{-1}\!$, $d=10\,\mu\mathrm{m}^2/\mathrm{s}$, $R=10\,\mu\mathrm{m}$, $\nu=2/3$ and $\lambda=50\,\mu\mathrm{m}/\mathrm{s}$.}
	\end{figure}
	
	Before computing the corresponding relaxation times of the system to steady-state from the Laplace space solutions in Eqs.~(\ref{eqn:hat-A-c}), (\ref{eqn:hat-A-n}), (\ref{eqn:hat-B-n}), (\ref{eqn:hat-I-n}) and (\ref{eqn:hat-I-c}), it is instructive to calculate the relaxation times for the fast diffusion case ($d\!\rightarrow\!\infty$), which leads to rapid homogenization of nuclear concentrations. In this limit of infinite diffusivity the associated equations take the form: 
	\begin{align}
		\label{eqn:fast-eq1}
		\frac{\partial\hspace{-1pt}\tilde{A}_c}{\partial t} &= \eta \tilde{I}_c - \frac{3\nu\lambda}{R}(\tilde{A}_c-\tilde{A}_n), \\[2pt]
		\frac{\partial\hspace{-1pt}\tilde{A}_n}{\partial t} &= -\alpha_0 \tilde{A}_n + \frac{3\lambda}{R}(\tilde{A}_c-\tilde{A}_n), \\[2pt]
		\frac{\partial\hspace{-1pt}\tilde{B}_n}{\partial t} &= -\beta_0 \tilde{B}_n + \alpha_0 \tilde{A}_n, \\[2pt]
		\frac{\partial\hspace{-0.5pt}\tilde{I}_n}{\partial t} &= \beta_0 \tilde{B}_n + \frac{3\lambda}{R}(\tilde{I}_c-\tilde{I}_n),\\[2pt]
		\frac{\partial\hspace{-0.5pt}\tilde{I}_c}{\partial t} &\label{eqn:fast-eq5}
		= -\eta \tilde{I}_c - \frac{3\nu\lambda}{R}(\tilde{I}_c-\tilde{I}_n),
	\end{align} where $\tilde{A}_c$ and $\tilde{A}_n$ denote the cytoplasmic and nuclear RanGDP concentrations; $\tilde{B}_n$ is the nuclear RanGTP concentration; while $\tilde{I}_c$ and $\tilde{I}_n$ denote the cytoplasmic and nuclear NTR--RanGTP concentrations. Notice that $\tilde{A}_n$, $\tilde{B}_n$ and $\tilde{I}_n$ are not functions of space here, representing mean concentrations defined over the whole nucleus. 	
		
	\begin{figure}[t!]\includegraphics[width=\columnwidth]{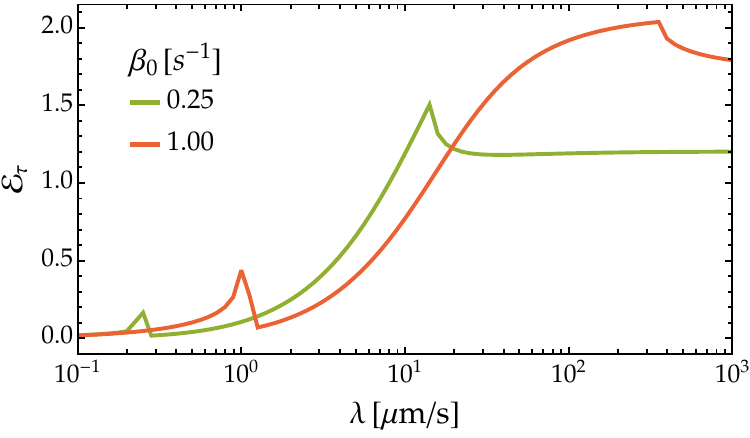}
		\caption{\label{fig:error-relaxation-times}  Relative increase of the dominant relaxation time $\tau$ compared to  $\tilde{\tau}$ in the fast diffusion case; namely,  $\mathcal{E}_\tau=\tau/\tilde{\tau}-1$.  We vary $\lambda$ at two values of $\beta_0$, with $\alpha_0 =5\,\mathrm{s}^{-1}\!$, $\eta =10\,\mathrm{s}^{-1}\!$, $\nu=2/3$, $d=10\,\mu\mathrm{m}^2/\mathrm{s}$, and $R=10\,\mu\mathrm{m}$.}
	\end{figure}
	
	This linear set of equations can be written as follows:
	\begin{equation}
		\frac{\partial\boldsymbol{\tilde{C}}}{\partial t} = \boldsymbol{M} \boldsymbol{\tilde{C}}
	\end{equation} where $\boldsymbol{\tilde{C}} = \hspace{-1pt}\big[\tilde{A}_c,\,\tilde{A}_n,\,\tilde{B}_n, \,\tilde{I}_n,\,\tilde{I}_c\hspace{1pt}\big]^{\mathsf{T}}$ and  $\boldsymbol{M}$ is given by
	\begin{equation}
		\boldsymbol{M}\hspace{-1pt}=\hspace{-2pt}
		\left[
		\begin{array}{ccccc}
			-\frac{3 \lambda  \nu }{R} & \frac{3 \lambda  \nu }{R} & 0 & 0 &
			\eta  \\[5pt]
			\frac{3 \lambda }{R} & -\alpha _0\hspace{-1pt}-\hspace{-1pt}\frac{3 \lambda }{R} & 0 & 0 & 0
			\\[5pt]
			0 & \alpha _0 & -\beta _0 & 0 & 0 \\[5pt]
			0 & 0 & \beta _0 & -\frac{3 \lambda }{R} & \frac{3 \lambda }{R} \\[5pt]
			0 & 0 & 0 & \frac{3 \lambda  \nu }{R} & -\eta\hspace{-1pt}-\hspace{-1pt}\frac{3 \lambda  \nu
			}{R} \\[5pt]
		\end{array}
		\right]\!.
	\end{equation} The relaxation times of the system to steady-state can be computed from the eigenvalues $\tilde{s}$ of matrix $\boldsymbol{M}$; namely, $\det(\boldsymbol{M}-\tilde{s}\boldsymbol{I}) = 0$, which shows that one of the eigenvalues is zero, while the other are given by roots of the following polynomial equation:  
	\begin{equation}
		\label{eqn:eigenvalues-fast-diffusion}
		\tilde{s}^4 + r_3\,\tilde{s}^3+r_2\,\tilde{s}^2+ r_1\,\tilde{s}+r_0 = 0, 
	\end{equation} where $r_j$ are constants ($j\in\{0,1,2,3\}$) that can be  computed in terms of $\alpha_0$, $\beta_0$, $\eta$, $\nu$, $\lambda$, and $R$. The real part of all non-zero eigenvalues are negative, from which the relaxation times $\tilde{\tau} = -1/\mathrm{Re}(\tilde{s})$ can be computed; as shown in Fig~2(a) and Fig.~\ref{fig:relaxation-times-fast-diffusion}. The imaginary part of the dominant eigenvalues can become nonzero as we vary the permeability $\lambda$, as shown in the inset plot of Fig.~\ref{fig:relaxation-times-fast-diffusion}. 
	
	In the limit of rapid GTP hydrolysis rate ($\eta\rightarrow\infty$) as well as fast nucleotide exchange ($\alpha_0\rightarrow\infty$), the system of Eqs.~(\ref{eqn:fast-eq1}--\ref{eqn:fast-eq5}) reduces to a simpler three-dimensional system in terms of only cytoplasmic RanGDP, nuclear RanGTP, and nuclear NTR--RanGTP; namely,
	\begin{align}
		\label{eqn:reduced-fast-eq1}
		\frac{\partial\hspace{-1pt}\tilde{A}_c}{\partial t} &= \frac{3\nu\lambda}{R}\tilde{I}_n - \frac{3\nu\lambda}{R}(\tilde{A}_c-\tilde{A}_n), \\[2pt]
		\frac{\partial\hspace{-1pt}\tilde{B}_n}{\partial t} &= -\beta_0 \tilde{B}_n + \frac{3\lambda}{R}(\tilde{A}_c-\tilde{A}_n), \\[2pt]
		\frac{\partial\hspace{-0.5pt}\tilde{I}_n}{\partial t} &\label{eqn:reduced-fast-eq3}
		= \beta_0 \tilde{B}_n - \frac{3\lambda}{R}\tilde{I}_n.
	\end{align} The characteristic polynomial equation in Eq.~(\ref{eqn:eigenvalues-fast-diffusion}) now reduces to
	\begin{equation}
		\label{eqn:eigenvalues-fast-diffusion-reduced}
		\tilde{s}^2\! +\tilde{s}\,\frac{3\lambda  (\nu\!+\!1)\!+\!\beta_0 R}{R}\!+\!\frac{3 \lambda\hspace{-1pt}\left[3 \nu \lambda\!+\!\beta_0 R\!\left(\nu\!+\!1\right) \right]}{R^2}\!= 0,
	\end{equation} which allows us to determine in exact form the solution of the eigenvalues $\tilde{s}$. The imaginary part of $\tilde{s}$ is nonzero when the discriminant of Eq.~(\ref{eqn:eigenvalues-fast-diffusion-reduced}) becomes negative, namely we require $\lambda$ to be in the range
	\begin{equation}
		\frac{\beta _0 R}{3 \left(\sqrt{\nu }-1\right)^2}>\lambda >\frac{\beta _0 R}{3
			\left(\sqrt{\nu }+1\right)^2},\;\,\text{with }\nu\neq1,
	\end{equation} or $\lambda>\beta_0R/12$ when ratio $\nu=1$. This shows that the onset of the oscillation occurs at higher values of $\lambda$ as we increase the rate $\beta_0$; see Fig.~\ref{fig:figure_2}(a).

	We now turn back to case when diffusivity $d$ is finite. To obtain the relaxation times in this case from the associated Laplace solutions in Eqs.~(\ref{eqn:hat-A-c}), (\ref{eqn:hat-A-n}), (\ref{eqn:hat-B-n}), (\ref{eqn:hat-I-n}) and (\ref{eqn:hat-I-c}), we need to locate the complex  poles of functions that have negative real part. Note that the pole at $s=0$ corresponds to the steady-state solutions found previously through the final value theorem.  Since the solutions of $\hat{A}_n$, $\hat{B}_n$, $\hat{I}_n$, and $\hat{I}_c$ are all linearly related to $\hat{A}_c$, then the position of their poles is the same for each of them. The dominant relaxation time $\tau$ is given by 
	\begin{equation}
		\tau = -1/\mathrm{Re}(s^{\star}),
	\end{equation} where $s^\star$ is the pole with the smallest and strictly negative real part, as shown in Fig.~\ref{fig:complex-plane}. By numerically tracking the position of the pole $s^\star$ as function of $\lambda$, the transition from an over-damped decay, with $\mathrm{Im}(s^{\star}) = 0$, to an under-damped decay, where $\mathrm{Im}(s^\star) \neq 0$, can be seen as we increase the permeability; see Fig.~\ref{fig:figure_2}(a). By comparing the dominant relaxation time $\tilde{\tau}$ in the fast diffusion case with $\tau$, we find that a significant relative increase, (defined by $\mathcal{E}_\tau = {\tau}/\tilde{\tau}-1$) at large values of the nuclear pore permeability $\lambda$, as shown in Fig.~\ref{fig:error-relaxation-times}.
	
	\section{Non-homogeneous RanGEF} 
	
	Here we consider that distribution of the RanGEF is not uniform within the nucleus, with the nucleotide exchange rate $\alpha(r)$ being a function of the radial distance. For simplicity, we  assume that $\alpha$ is a piecewise constant function. We consider two extreme cases; first, all of the RanGEF is distributed within a spherical ball of radius $R\hspace{1pt}\varepsilon$ at the center of the nucleus; and second, all of the RanGEF is localized in a spherical shell at the nuclear envelope of thickness $R\hspace{1pt}(1-\delta)$. We study how the steady-state concentrations of each molecular species, as well as their corresponding total number, is affected by the non-homogeneous distribution of RanGEF.\\
	\indent 
	In the case where all of the RanGEF is located in a spherical ball at the center of the nucleus, the nucleotide exchange rate for Ran is a piecewise constant function: 
	\begin{equation}
		\alpha_\mathrm{ball}(r) = \begin{cases} \,a_0 F_\mathrm{ball} & 0<r\leq\varepsilon, \\[2pt] \,0 & \varepsilon<r< 1,\end{cases}
	\end{equation} where $a_0$ is a constant, and $F_\mathrm{ball}$ is the RanGEF concentration within the ball. Thus, we need to solve the nuclear equations (\ref{eqn:dyn-A-linear}), (\ref{eqn:dyn-B-linear}) and (\ref{eqn:dyn-I-linear}) on two separate domains, using matching boundary conditions at $r=\varepsilon$; namely, we demand that concentration and its derivative are continuous at $r=\varepsilon$.\\
	\indent 
	Similarly, in the case where the RanGEF is distributed only in a shell near the nuclear boundary, the nucleotide exchange rate is given by  
	\begin{equation}
		\alpha_\mathrm{shell}(r) = \begin{cases} \,0 & 0<r\leq 1-\delta, \\[2pt] \,a_0 F_\mathrm{shell} & 1-\delta<r< 1,\end{cases}
	\end{equation} where $F_\mathrm{shell}$ is the RanGEF concentration in the shell. Thus, a continuity boundary condition is imposed at the radial position $r=1-\delta$ where the concentrations and their derivatives are continuous.\\
	\indent  
	To directly compare with the previous case of homogeneous RanGEF throughout the whole nucleus, we demand that the RanGEF number is the same in both cases: 
	$\frac{4}{3}\pi R^3(1 - \delta^3)F_\mathrm{shell} = \frac{4}{3}\pi R^3 F_0$, which gives
	\begin{equation}
		\alpha_\mathrm{shell}(r) = \begin{cases} \,0 & 0<r\leq 1-\delta, \\[2pt] \,\alpha_0/(1-\delta^3) & 1-\delta<r< 1,\end{cases}
	\end{equation} where $\alpha_0=a_0 F_0$, being the same as the rate defined previously in Appendix~A. Similarly, we demand the number of RanGEF to be the same in the ball as in the entirely uniform case:  $\frac{4}{3}\pi R^3\varepsilon^3 F_\mathrm{ball} = \frac{4}{3}\pi R^3 F_0$. Hence,
	\begin{equation}
		\alpha_\mathrm{ball}(r) = \begin{cases} \,\alpha_0 /\varepsilon^3 & 0<r\leq\varepsilon, \\[2pt] \,0 & \varepsilon<r< 1.\end{cases}
	\end{equation}
	Given these two static distributions for RanGEF, the steady-state solutions for the concentrations of RanGDP, RanGTP, and NTR--RanGTP can be determined in exact form and compared against the entirely uniform case; see concentration profiles in Fig.~\ref{fig:figure_2}(b).
	\subsection{Steady-state profiles --- the shell case}
	Using Eq.~(\ref{eqn:dyn-A-linear}), the steady-state equation for nuclear RanGDP can be written as follows:
	\begin{equation}
		\begin{cases} 
			\frac{1}{r^2}\frac{\partial}{\partial r}\!\left[r^2\frac{\partial \bar{A}^{\mathrm{so}}_n}{\partial r}\right]\!=0,
			& 0<r\leq1\!-\!\delta, \\[8pt] 
			\frac{1}{r^2}\frac{\partial}{\partial r}\!\left[r^2\frac{\partial \bar{A}^{\mathrm{si}}_n}{\partial r}\right]\!= k^{\hspace{0.5pt}\mathrm{s}}_{\alpha}  \bar{A}^{\mathrm{si}}_n,
			& 1\!-\!\delta<r\leq 1.\end{cases}
	\end{equation} where $k^{\hspace{0.5pt}\mathrm{s}}_{\alpha} = R^2\alpha_0/[d(1-\delta^3)]$, while $\bar{A}_n^{\mathrm{si}}$ and $\bar{A}_n^{\mathrm{so}}$ denote the concentrations inside and outside the shell, respectively. The nontrivial solution of $\bar{A}_n^{\mathrm{si}}$ is found to be
	\begin{equation}
		\bar{A}_n^{\mathrm{si}}(r) = \frac{c^{\hspace{0.5pt}\mathrm{s}}_1 \cosh(r\sqrt{k^{\hspace{0.5pt}\mathrm{s}}_\alpha})+c^{\hspace{0.5pt}\mathrm{s}}_2 \sinh(r\sqrt{k^{\hspace{0.5pt}\mathrm{s}}_\alpha})}{r},
	\end{equation} while $\bar{A}_n^{\mathrm{so}} = c^{\hspace{0.5pt}\mathrm{s}}_3$, where $c^{\hspace{0.5pt}\mathrm{s}}_1$, $c^{\hspace{0.5pt}\mathrm{s}}_2$ and $c^{\,\mathrm{s}}_3$ are constants to be determined from the boundary conditions; namely,
	\begin{align}
		&\bar{A}_n^{\mathrm{si}}(1-\delta) = \bar{A}_n^{\mathrm{so}}(1-\delta),\\[5pt]	
		&\frac{\partial\hspace{-1pt}\bar{A}_n^{\mathrm{si}}}{\partial r}(1-\delta) = 	\frac{\partial\hspace{-1pt}\bar{A}_n^{\mathrm{so}}}{\partial r}(1-\delta),\\[5pt]	
		& \frac{\partial\hspace{-1pt}\bar{A}^{\mathrm{si}}_n}{\partial r}(1) = \Lambda\! \left[\bar{A}^{\mathrm{s}}_{c} - \bar{A}^{\mathrm{si}}_n(1)\right]\!,
	\end{align} where $\bar{A}^{\mathrm{s}}_{c}$ is the cytoplasmic concentration of RanGDP at steady-state for the shell case. 
	
	By Eq.~(\ref{eqn:dyn-B-linear}), the steady-state equation for nuclear concentration of RanGTP is given by
	\begin{equation}
		\begin{cases} 
			\frac{1}{r^2}\frac{\partial}{\partial r}\!\left[r^2\frac{\partial \bar{B}^{\mathrm{so}}_n}{\partial r}\right]\!=k_\beta \bar{B}^{\mathrm{so}}_n,
			& 0<r\leq1\!-\!\delta, \\[8pt] 
			\frac{1}{r^2}\frac{\partial}{\partial r}\!\left[r^2\frac{\partial \bar{B}^{\mathrm{si}}_n}{\partial r}\right]\!= k_\beta \bar{B}^{\mathrm{si}}_n\!-k^{\,\mathrm{s}}_{\alpha}  \bar{A}^{\mathrm{si}}_n,
			& 1\!-\!\delta<r\leq 1,\end{cases}
	\end{equation} where $k_{\beta} = R^2\beta_0/d$. Similarly, $\bar{B}^{\mathrm{si}}_n$ and $\bar{B}^{\mathrm{so}}_n$ denote the concentrations of RanGTP inside and outside the shell, respectively. The solution of $\bar{B}^{\mathrm{si}}_n$ is given by
	\begin{align}
		\bar{B}^{\mathrm{si}}_n(r) &=  \frac{c^{\,\mathrm{s}}_1 k_\alpha^{\,\mathrm{s}}\cosh(r\sqrt{k^{\,\mathrm{s}}_\alpha})+c^{\,\mathrm{s}}_2k_\alpha^{\,\mathrm{s}} \sinh(r\sqrt{k^{\,\mathrm{s}}_\alpha})}{r(k_\beta-k^{\,\mathrm{s}}_\alpha)}\notag\\[6pt] 
		&\quad+\frac{c^{\,\mathrm{s}}_4\cosh(r\sqrt{k^{\,\mathrm{s}}_\alpha})+c^{\,\mathrm{s}}_5 \sinh(r\sqrt{k^{\,\mathrm{s}}_\alpha})}{r},
	\end{align} while the solution of $\bar{B}^{\mathrm{so}}_n$ is found to be
	\begin{align}
		\bar{B}^{\mathrm{so}}_n(r) = \frac{c^{\,\mathrm{s}}_6\sinh(r\sqrt{k_\beta})}{r},
	\end{align} where $c^{\,\mathrm{s}}_4$, $c^{\,\mathrm{s}}_5$ and $c^{\,\mathrm{s}}_6$ are constants to be determined from
	\begin{align}
		&\bar{B}_n^{\mathrm{si}}(1-\delta) = \bar{B}_n^{\mathrm{so}}(1-\delta),\\[5pt]	
		&\frac{\partial\hspace{-1pt}\bar{B}_n^{\mathrm{si}}}{\partial r}(1-\delta) = 	\frac{\partial\hspace{-1pt}\bar{B}_n^{\mathrm{so}}}{\partial r}(1-\delta),\\[5pt]	
		& \frac{\partial\hspace{-1pt}\bar{B}^{\mathrm{si}}_n}{\partial r}(1) = 0.
	\end{align}
	
		\begin{figure*}[t!]\includegraphics[width=\textwidth]{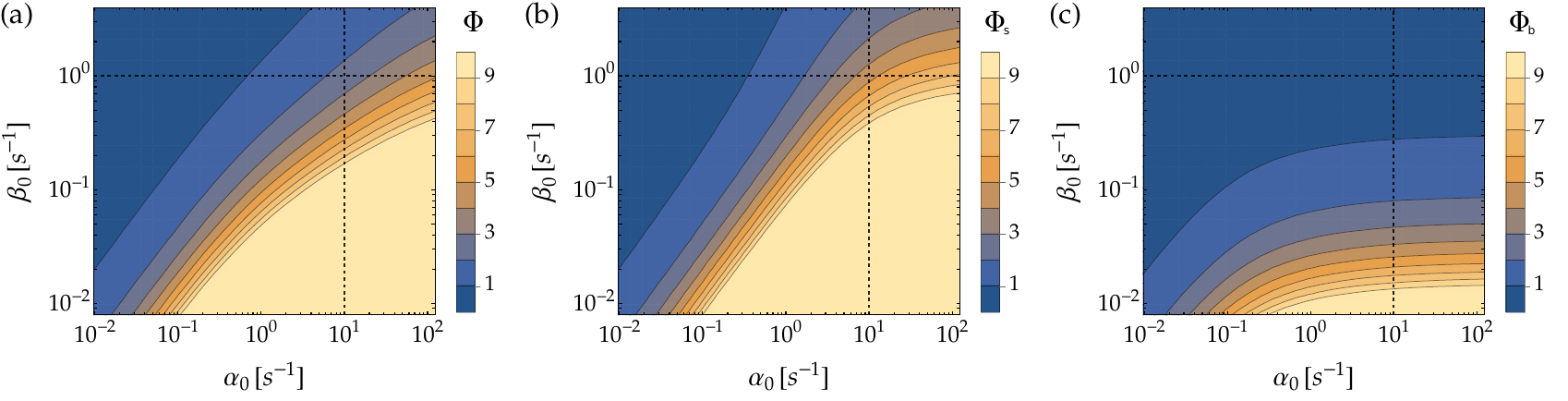}
		\caption{\label{fig:grid-Phi} (a) Steady-state nuclear-to-cytoplasm ratio of Ran molecules as a function of $\alpha_0$ and $\beta_0$ for the case where RanGEF is uniformly distributed  throughout the whole nucleus. (b) Steady-state ratio $\Phi_\mathrm{s}$ in the case where RanGEF is only localized in a spherical shell, with $\delta=0.98$. Here, the total number of RanGEF molecules is the same as in the uniform case of subfigure~(a). (c)~Steady-state $\Phi_\mathrm{b}$ for the case of RanGEF distributed in a spherical ball, with $\varepsilon=(1-\delta^3)^{-1/3}\approx0.39$, which enforces that we have the same number of RanGEF molecules as in (a) and (b). For all of the sub-figures, we set $\eta =10\,\mathrm{s}^{-1}\!$, $d=10\,\mu\mathrm{m}^2/\mathrm{s}$, $R=10\,\mu\mathrm{m}$, $\nu=2/3$ and $\lambda=50\,\mu\mathrm{m}/\mathrm{s}$. The black dashed lines highlight the same parameter point in each of the plots.
		}
	\end{figure*}

		\begin{figure*}[t!]
	\centering
	\includegraphics[width=\textwidth]{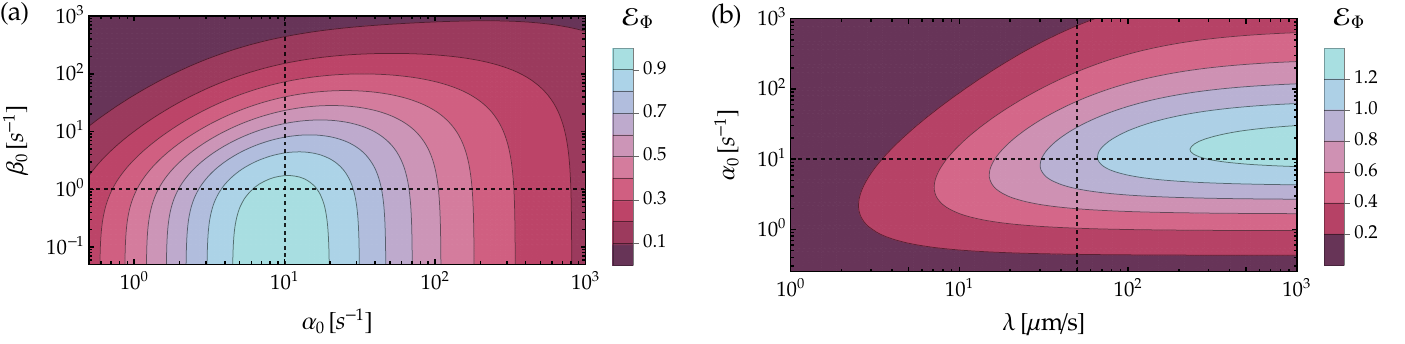}
	\caption{\label{fig:plotErrs} Relative increases $\mathcal{E}_\Phi$ in the ratio $\Phi_\mathrm{b}(\delta)$, with RanGEF distributed only in a spherical shell of thickness $R(1-\delta)$, compared to the entirely uniform case ($\delta\rightarrow1$), {\it i.e.}~$\mathcal{E}_\Phi = \Phi_\mathrm{b}(\delta)/\Phi_\mathrm{b}(\delta\!\rightarrow\!1)\!-\!1$. In subfigure (a), we plot $\mathcal{E}_\Phi$ as function of $\alpha_0$ and $\beta_0$ at fixed $\lambda=50\,\mu\mathrm{m}/\mathrm{s}$, while in subfigure (b), we vary $\alpha_0$ and $\lambda$ at fixed $\beta_0=1\,\mathrm{s}^{-1}$. Here, $\delta=0.98$, $\nu=2/3$, $\eta =10\,\mathrm{s}^{-1}\!$, $d=10\,\mu\mathrm{m}^2/\mathrm{s}$, and $R=10\,\mu\mathrm{m}$. Dashed lines show the same parameter point in (a) and (b).
	}
\end{figure*}

	Using Eq.~(\ref{eqn:dyn-I-linear}), the steady-state equation for nuclear NTR--RanGTP concentration becomes
	\begin{equation}
		\begin{cases} 
			\frac{1}{r^2}\frac{\partial}{\partial r}\!\left[r^2\frac{\partial \bar{I}^{\mathrm{so}}_n}{\partial r}\right]\!= -k_\beta \bar{B}^{\mathrm{so}}_n,
			& 0<r\leq1\!-\!\delta, \\[8pt] 
			\frac{1}{r^2}\frac{\partial}{\partial r}\!\left[r^2\frac{\partial \bar{I}^{\mathrm{si}}_n}{\partial r}\right]\!= -k_\beta \bar{B}^{\mathrm{si}}_n,
			& 1\!-\!\delta<r\leq 1,\end{cases}
	\end{equation} where $\bar{I}^{\mathrm{si}}_n$ and $\bar{I}^{\mathrm{so}}_n$ are the nuclear NTR--RanGTP concentrations inside and outside the shell. We find
	\begin{align}
		\!\bar{I}^{\mathrm{si}}_n(r) &=  \frac{c^{\,\mathrm{s}}_7}{r}+c^{\,\mathrm{s}}_8 -\frac{c^{\,\mathrm{s}}_4\cosh(r\sqrt{k^{\,\mathrm{s}}_\alpha})+c^{\,\mathrm{s}}_5 \sinh(r\sqrt{k^{\,\mathrm{s}}_\alpha})}{r} \notag\\[6pt] 
		&\quad+\frac{c^{\,\mathrm{s}}_1 k_\beta\cosh(r\sqrt{k^{\,\mathrm{s}}_\alpha})+c^{\,\mathrm{s}}_2 k_\beta \sinh(r\sqrt{k^{\,\mathrm{s}}_\alpha})}{r(k^{\,\mathrm{s}}_\alpha-k_\beta)},
	\end{align} while the outer solution $\bar{I}^{\mathrm{so}}_n$ is given by
	\begin{equation}
		\bar{I}^{\mathrm{so}}_n(r) = c^{\,\mathrm{s}}_9 - \frac{c^{\,\mathrm{s}}_6\sinh(r\sqrt{k_\beta})}{r},
	\end{equation} where the constants $c^{\,\mathrm{s}}_7$, $c^{\,\mathrm{s}}_8$ and $c^{\,\mathrm{s}}_9$ are found by imposing:
	\begin{align}
		&\bar{I}_n^{\mathrm{si}}(1-\delta) = \bar{I}_n^{\mathrm{so}}(1-\delta),\\[5pt]	
		&\frac{\partial\hspace{-1pt}\bar{I}_n^{\mathrm{si}}}{\partial r}(1-\delta) = 	\frac{\partial\hspace{-1pt}\bar{I}_n^{\mathrm{so}}}{\partial r}(1-\delta),\\[5pt]	
		& \frac{\partial\hspace{-1pt}\bar{I}^{\mathrm{si}}_n}{\partial r}(1) = \Lambda \!\left[\bar{I}^{\mathrm{s}}_{c} - \bar{I}^{\mathrm{si}}_n(1)\right]\!,
	\end{align} where $\bar{I}^{\mathrm{s}}_n$ is the cytoplasmic concentration at steady-state of NTR--RanGTP molecules. From Eq.~(\ref{eqn:dyn-Ic-linear}), the steady-state value of $\bar{I}^{\mathrm{s}}_n$ can be written as follows:
	\begin{equation}
		\bar{I}^{\mathrm{s}}_{c}= \frac{3\nu \Lambda}{k_\eta + 3\nu \Lambda}\,\bar{I}^{\mathrm{si}}_n(1).
	\end{equation} This allows us to find the constants $c^{\,\mathrm{s}}_j$, $j\in\{1,2,\dots,9\}$, in terms of the unknown steady-state value of $\bar{A}^{\mathrm{s}}_c$, and the dimensionless parameters $k_\alpha^{\hspace{0.5pt}\mathrm{s}}$, $k_\beta$, $k_\eta$, $\nu$, $\Lambda$, and $\delta$. The steady-state $\bar{A}^{\mathrm{s}}_c$ is found  by imposing the conservation of the net number of molecules in the cell, as in Eq.~(\ref{eqn:conservation-2}).

\begin{figure*}[t!]
	\centering
	\includegraphics[width=\textwidth]{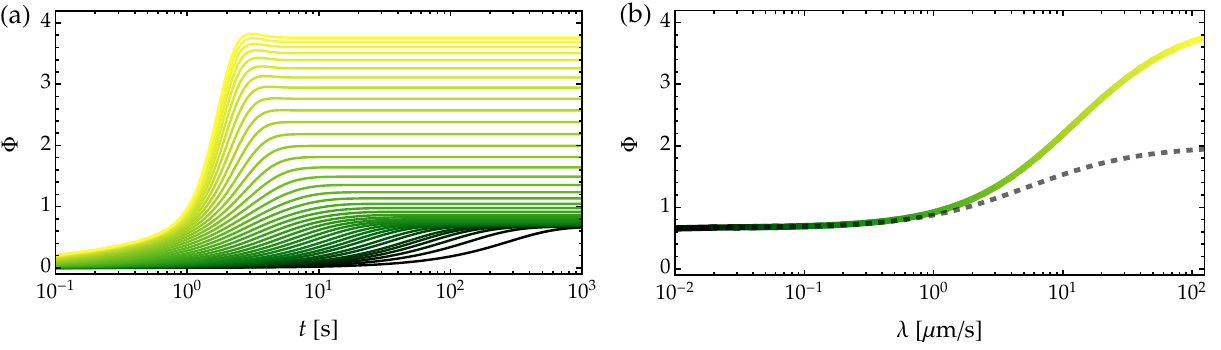}
	\caption{\label{fig:plot_phi_vs_lambda} (a) The nuclear-to-cytoplasmic ratio $\Phi$ for various values of the permeability $\lambda$; all simulations start with the same initial conditions. (b) The long-time value of $\Phi$ versus $\lambda$; same color convention as in (a). Dashed curve is the steady-state value of $\Phi$ when RanGEF is uniformly distributed in the nucleus, with concentration $F_0$ and same number of RanGEF molecules as that of the solid line.  Here, $\beta_0 = 1\,\mathrm{s}^{-1}$, $\alpha_0 = 10\,\mathrm{s}^{-1}$, $\eta =10\,\mathrm{s}^{-1}\!$, $d=10\,\mu\mathrm{m}^2/\mathrm{s}$, $R=10\,\mu\mathrm{m}$, $\nu=2/3$, and $\gamma F_0 = 50\,\mathrm{s}^{-1}$.
	}
\end{figure*}

	\subsection{Steady-state profiles --- the ball case}

	From Eq.~(\ref{eqn:dyn-A-linear}), the steady-state equation for nuclear RanGDP is given by
	\begin{equation}
		\begin{cases} 
			\frac{1}{r^2}\frac{\partial}{\partial r}\!\left[r^2\frac{\partial \bar{A}^{\mathrm{bo}}_n}{\partial r}\right]\!=0,
			& \varepsilon<r\leq1, \\[8pt] 
			\frac{1}{r^2}\frac{\partial}{\partial r}\!\left[r^2\frac{\partial \bar{A}^{\mathrm{bi}}_n}{\partial r}\right]\!= k^{\hspace{0.5pt}\mathrm{b}}_{\alpha}  \bar{A}^{\mathrm{bi}}_n,
			& 0<r\leq\varepsilon.\end{cases}
	\end{equation} where $k^{\hspace{0.5pt}\mathrm{b}}_{\alpha} = R^2\alpha_0/(d\hspace{0.5pt}\varepsilon^3)$, whilst $\bar{A}_n^{\mathrm{bi}}$ and $\bar{A}_n^{\mathrm{bo}}$ are the concentrations inside and outside the spherical ball, respectively. The solution of $\bar{A}_n^{\mathrm{bi}}$ is found to be
	\begin{equation}
		\bar{A}_n^{\mathrm{bi}}(r) = \frac{c^{\hspace{0.5pt}\mathrm{b}}_1 \sinh(r\sqrt{k^{\hspace{0.5pt}\mathrm{b}}_\alpha})}{r},
	\end{equation} while the solution of $\bar{A}_n^{\mathrm{bo}}$ is given by
	\begin{equation}
		\bar{A}_n^{\mathrm{bo}}(r) = \frac{c^{\hspace{0.5pt}\mathrm{b}}_2}{r}+ c^{\hspace{0.5pt}\mathrm{b}}_3,
	\end{equation} where $c^{\hspace{0.5pt}\mathrm{b}}_1$, $c^{\hspace{0.5pt}\mathrm{b}}_2$ and $c^{\,\mathrm{b}}_3$ are constants to be determined from:
	\begin{align}
		&\bar{A}_n^{\mathrm{bi}}(\varepsilon) = \bar{A}_n^{\mathrm{bo}}(\varepsilon),\\[5pt]	
		&\frac{\partial\hspace{-1pt}\bar{A}_n^{\mathrm{bi}}}{\partial r}(\varepsilon) = 	\frac{\partial\hspace{-1pt}\bar{A}_n^{\mathrm{bo}}}{\partial r}(\varepsilon),\\[5pt]	
		& \frac{\partial\hspace{-1pt}\bar{A}^{\mathrm{bo}}_n}{\partial r}(1) = \Lambda\! \left[\bar{A}^{\mathrm{b}}_{c} - \bar{A}^{\mathrm{bo}}_n(1)\right]\!,
	\end{align} where $\bar{A}^{\mathrm{b}}_{c}$ is the steady-state value of the cytoplasmic concentration of RanGDP in the ball case.

	By Eq.~(\ref{eqn:dyn-B-linear}), the steady-state equation for nuclear concentration of RanGTP is given by
	\begin{equation}
		\begin{cases} 
			\frac{1}{r^2}\frac{\partial}{\partial r}\!\left[r^2\frac{\partial \bar{B}^{\mathrm{bo}}_n}{\partial r}\right]\!=k_\beta \bar{B}^{\mathrm{bo}}_n,
			& \varepsilon<r\leq1, \\[8pt] 
			\frac{1}{r^2}\frac{\partial}{\partial r}\!\left[r^2\frac{\partial \bar{B}^{\mathrm{bi}}_n}{\partial r}\right]\!= k_\beta \bar{B}^{\mathrm{bi}}_n\!-k^{\,\mathrm{b}}_{\alpha}  \bar{A}^{\mathrm{bi}}_n,
			& 0<r\leq\varepsilon,\end{cases}
	\end{equation} where $\bar{B}^{\mathrm{bi}}_n$ and $\bar{B}^{\mathrm{bo}}_n$ are the concentrations of RanGTP inside and outside the ball, respectively. We find that
	\begin{align}
		\bar{B}^{\mathrm{bi}}_n(r) &=  \frac{c^{\hspace{0.5pt}\mathrm{b}}_4\sinh(r\sqrt{k_\beta})}{r}+\frac{c^{\hspace{0.5pt}\mathrm{b}}_1 k^{\hspace{0.5pt}\mathrm{b}}_\alpha \sinh(r\sqrt{k^{\hspace{0.5pt}\mathrm{b}}_\alpha})}{r(k_\beta-k^{\hspace{0.5pt}\mathrm{b}}_\alpha)},\\[5pt]
		\bar{B}^{\mathrm{so}}_n(r) &= \frac{c^{\,\mathrm{b}}_5\cosh(r\sqrt{k_\beta})+c^{\,\mathrm{b}}_6\sinh(r\sqrt{k_\beta})}{r},
	\end{align} where $c^{\,\mathrm{b}}_4$, $c^{\,\mathrm{b}}_5$ and $c^{\,\mathrm{b}}_6$ are constants to be determined from
	\begin{align}
		&\bar{B}_n^{\mathrm{bi}}(\varepsilon) = \bar{B}_n^{\mathrm{bo}}(\varepsilon),\\[5pt]	
		&\frac{\partial\hspace{-1pt}\bar{B}_n^{\mathrm{bi}}}{\partial r}(\varepsilon) = 	\frac{\partial\hspace{-1pt}\bar{B}_n^{\mathrm{bo}}}{\partial r}(\varepsilon),\\[5pt]	
		& \frac{\partial\hspace{-1pt}\bar{B}^{\mathrm{bo}}_n}{\partial r}(1) = 0.
	\end{align}
	
	Using Eq.~(\ref{eqn:dyn-I-linear}), we write 
	\begin{equation}
		\begin{cases} 
			\frac{1}{r^2}\frac{\partial}{\partial r}\!\left[r^2\frac{\partial \bar{I}^{\mathrm{bo}}_n}{\partial r}\right]\!= -k_\beta \bar{B}^{\mathrm{bo}}_n,
			& \varepsilon<r\leq1, \\[8pt] 
			\frac{1}{r^2}\frac{\partial}{\partial r}\!\left[r^2\frac{\partial \bar{I}^{\mathrm{bi}}_n}{\partial r}\right]\!= -k_\beta \bar{B}^{\mathrm{bi}}_n,
			& 0<r\leq\varepsilon,\end{cases}
	\end{equation} where $\bar{I}^{\mathrm{bi}}_n$ and $\bar{I}^{\mathrm{bo}}_n$ are nuclear NTR--RanGTP concentrations inside and outside the ball. Their solutions are
	\begin{align}
		\!\bar{I}^{\mathrm{bi}}_n\!&=  c^{\,\mathrm{b}}_7 -\frac{c^{\hspace{0.5pt}\mathrm{b}}_4\hspace{-1pt}\sinh(r\sqrt{k_\beta})}{r}+\frac{c^{\hspace{0.5pt}\mathrm{b}}_1 k_\beta \hspace{-1pt}\sinh(r\sqrt{k^{\hspace{0.5pt}\mathrm{b}}_\alpha})}{r(k^{\hspace{0.5pt}\mathrm{b}}_\alpha-k_\beta)},\\[5pt]
		\bar{I}^{\mathrm{so}}_n\!&=  c^{\,\mathrm{b}}_9\! -\!\frac{c^{\,\mathrm{b}}_8\! +\!c^{\,\mathrm{b}}_5\hspace{-1pt}\cosh(r\sqrt{k_\beta})\!+\!c^{\,\mathrm{b}}_6\hspace{-1pt}\sinh(r\sqrt{k_\beta})}{r},
	\end{align} where $c^{\,\mathrm{b}}_7$, $c^{\,\mathrm{b}}_8$ and $c^{\,\mathrm{b}}_9$ are found from the conditions:
	\begin{align}
		&\bar{I}_n^{\mathrm{bi}}(\varepsilon) = \bar{I}_n^{\mathrm{bo}}(\varepsilon),\\[5pt]	
		&\frac{\partial\hspace{-1pt}\bar{I}_n^{\mathrm{bi}}}{\partial r}(\varepsilon) = 	\frac{\partial\hspace{-1pt}\bar{I}_n^{\mathrm{bo}}}{\partial r}(\varepsilon),\\[5pt]	
		& \frac{\partial\hspace{-1pt}\bar{I}^{\mathrm{bo}}_n}{\partial r}(1) = \Lambda \!\left[\bar{I}^{\mathrm{b}}_{c} - \bar{I}^{\mathrm{bo}}_n(1)\right]\!,
	\end{align} where $\bar{I}^{\mathrm{b}}_{c}$ is the steady-state cytoplasmic concentration of NTR--RanGTP complexes, which via Eq.~(\ref{eqn:dyn-Ic-linear}) becomes
	\begin{equation}
		\bar{I}^{\mathrm{b}}_{c}= \frac{3\nu \Lambda}{k_\eta + 3\nu \Lambda}\,\bar{I}^{\mathrm{bo}}_n(1).
	\end{equation} This allows us to determine $c^{\,\mathrm{b}}_j$, $j\in\{1,2,\dots,9\}$, in terms of $\bar{A}^{\mathrm{b}}_c$. The latter is found  by imposing the conservation of the net number of molecules in the cell, as in Eq.~(\ref{eqn:conservation-2}). 
	
	\subsection{Net number of molecules and their ratios}  
	
	\begin{figure}[t!]
	\centering
	\includegraphics[width=\columnwidth]{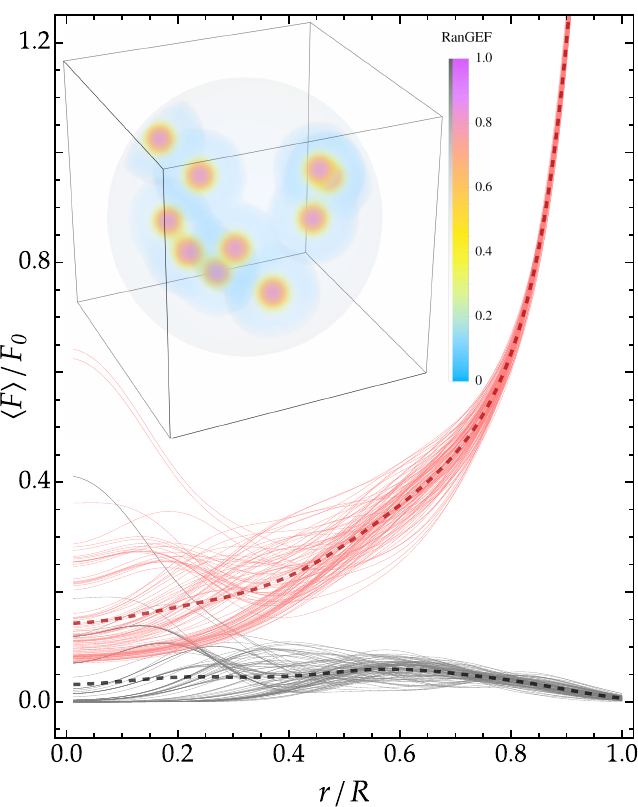}
	\caption{\label{fig:plot_F_init_vs_F_final} The initial (black curves) and long-time (red curves) of the angular average $\langle F\rangle$ as a function of the radial distance for 100 different random initial realizations of RanGEF concentration. The initial distributions are chosen as 10 Gaussian clusters of equal standard deviation whose mean is randomly placed within the nucleus. An example of initial data of RanGEF is shown in the inset plot, where each color is associated with an opacity level, indicated by the gray level in the color bar. Dashed curves are the corresponding averages over the all the initial (black) and long--time (red) profiles. Here, $\beta_0 = 1\,\mathrm{s}^{-1}$, $\alpha_0 = 10\,\mathrm{s}^{-1}$, $\eta =10\,\mathrm{s}^{-1}\!$, $d=10\,\mu\mathrm{m}^2/\mathrm{s}$, $R=10\,\mu\mathrm{m}$, $\nu=2/3$, $\lambda = 50\,\mu\mathrm{m}/\mathrm{s}$, and $\gamma F_0 = 50\,\mathrm{s}^{-1}$.
	}
	\end{figure}
	
	\begin{figure}[t!]
	\centering
	\includegraphics[width=\columnwidth]{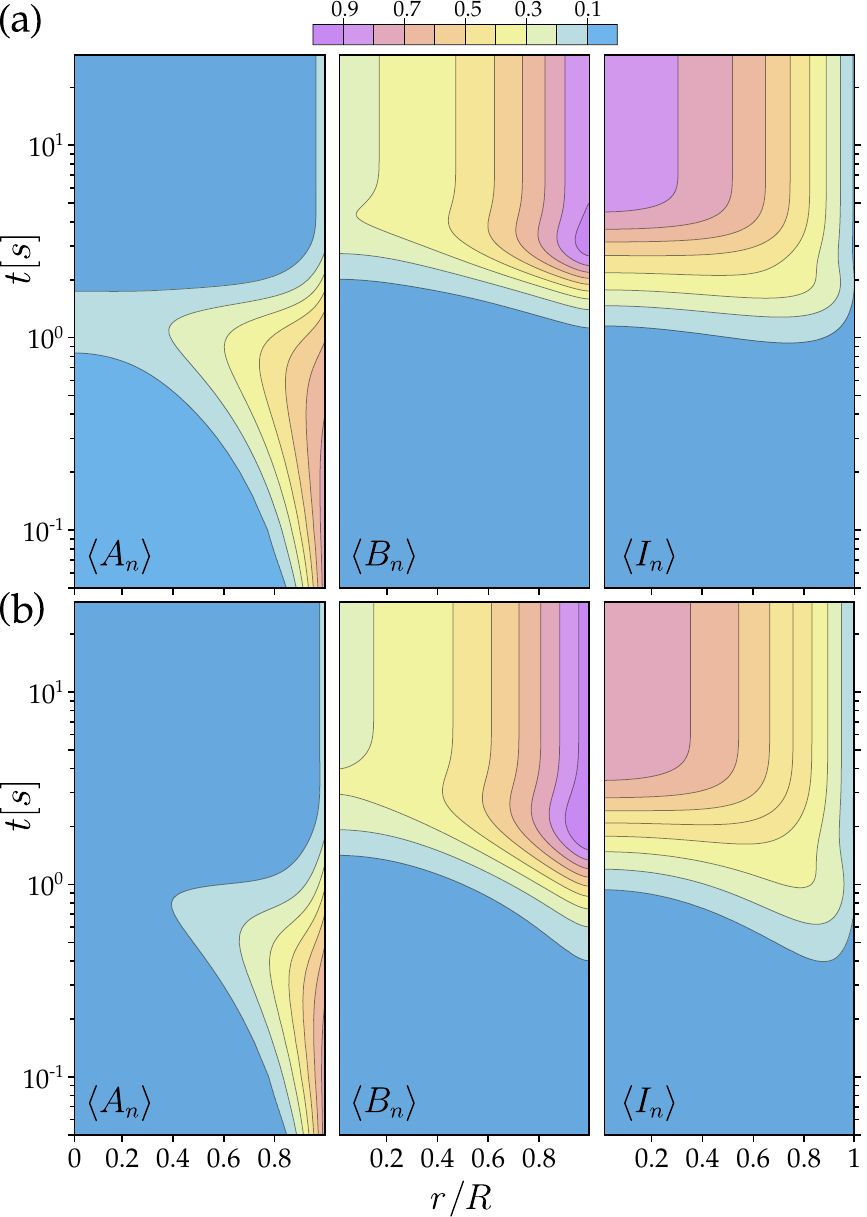}
	\caption{\label{fig:grid-plots-abi} Angular average of the nuclear RanGDP $\langle A_n\rangle$, RanGTP $\langle B_n\rangle$, and NTR--RanGTP $\langle I_n\rangle$ as a function of time and radial distance from the center of the nucleus. In subfigure (a), the initial RanGEF profile is uniform, representing only 10\% of the total number of RanGEF in the cell, while the other 90\% is in the cytoplasm as NTR--RanGEF. In subfigure (b), the initial distribution of RanGEF is given by random Gaussian clusters as shown in the inset of Fig.~10.  The remaining initial RanGEF is in the cytoplasm as NTR--RanGEF. For both cases $\beta_0 = 1\,\mathrm{s}^{-1}$, $\alpha_0 = 10\,\mathrm{s}^{-1}$, $\eta =10\,\mathrm{s}^{-1}\!$, $d=10\,\mu\mathrm{m}^2/\mathrm{s}$, $R=10\,\mu\mathrm{m}$, $\nu=2/3$, and $\gamma F_0 = 50\,\mathrm{s}^{-1}$.
	}
	\end{figure}
	By integrating the concentration fields over the nuclear volume, we can compute the respective net number of molecules. Using Eq.~(\ref{eqn:def-n-A}), total number of nuclear RanGDP at steady-state for the ball and shell cases is 
	\begin{align}
		\frac{\bar{n}_A^\mathrm{b}}{4\pi R^3} &= \int_{0}^{\varepsilon}\!\bar{A}_n^{\mathrm{bi}}(r)\,r^2\mathrm{d}r + \!\int_{\varepsilon}^{1}\!\bar{A}_n^{\mathrm{bo}}(r)\,r^2\mathrm{d}r,\\[3pt]
		\frac{\bar{n}_A^\mathrm{s}}{4\pi R^3} &= \int_{0}^{1-\delta}\!\bar{A}_n^{\mathrm{so}}(r)\,r^2\mathrm{d}r + \!\int_{1-\delta}^{1}\!\bar{A}_n^{\mathrm{si}}(r)\,r^2\mathrm{d}r,
	\end{align}
	Similarly, using Eq.~(\ref{eqn:def-n-A}), we have that
	\begin{align}
		\frac{\bar{n}_B^\mathrm{b}}{4\pi R^3} &= \int_{0}^{\varepsilon}\!\bar{B}_n^{\mathrm{bi}}(r)\,r^2\mathrm{d}r + \!\int_{\varepsilon}^{1}\!\bar{B}_n^{\mathrm{bo}}(r)\,r^2\mathrm{d}r,\\[3pt]
		\frac{\bar{n}_B^\mathrm{s}}{4\pi R^3} &= \int_{0}^{1-\delta}\!\bar{B}_n^{\mathrm{so}}(r)\,r^2\mathrm{d}r + \!\int_{1-\delta}^{1}\!\bar{B}_n^{\mathrm{si}}(r)\,r^2\mathrm{d}r,
	\end{align} where $\bar{n}_B^\mathrm{b}$ and $\bar{n}_B^\mathrm{s}$ is the net number of nuclear RanGTP in the ball and shell cases, respectively. From Eq.~(\ref{eqn:def-n-I}),
	\begin{align}
		\frac{\bar{n}_I^\mathrm{b}}{4\pi R^3} &= \int_{0}^{\varepsilon}\!\bar{I}_n^{\mathrm{bi}}(r)\,r^2\mathrm{d}r + \!\int_{\varepsilon}^{1}\!\bar{I}_n^{\mathrm{bo}}(r)\,r^2\mathrm{d}r,\\[3pt]
		\frac{\bar{n}_I^\mathrm{s}}{4\pi R^3} &= \int_{0}^{1-\delta}\!\bar{I}_n^{\mathrm{so}}(r)\,r^2\mathrm{d}r + \!\int_{1-\delta}^{1}\!\bar{I}_n^{\mathrm{si}}(r)\,r^2\mathrm{d}r,
	\end{align} where $\bar{n}_I^\mathrm{b}$ and $\bar{n}_I^\mathrm{s}$ is the net number of nuclear NTR--RanGTP in the ball and shell cases, respectively.
	
	By using the definition in Eq.~(\ref{eqn:Phi-def}) of the nuclear-to-cytoplasmic ratio of Ran molecules, its steady-state value in the ball and shell cases can be written as 
	\begin{align}
		\!\bar{\Phi}_\mathrm{b}\! = \frac{\bar{n}_A^\mathrm{b}+\bar{n}_B^\mathrm{b}+\bar{n}_I^\mathrm{b}}{\left(\bar{A}^{\mathrm{b}}_c+\bar{I}^{\mathrm{b}}_c\right)\!\mathcal{V}_c}\;\;\text{and}\;\; \bar{\Phi}_\mathrm{s}\! = \frac{\bar{n}_A^\mathrm{s}+\bar{n}_B^\mathrm{s}+\bar{n}_I^\mathrm{s}}{\left(\bar{A}^{\mathrm{s}}_c+\bar{I}^{\mathrm{s}}_c\right)\!\mathcal{V}_c},
	\end{align} respectively. These ratios are plotted in Fig.~\ref{fig:grid-Phi} as function of the rates $\alpha_0$ and $\beta_0$; see Fig.~\ref{fig:figure_2}(d). The relative increase in the steady-state value of $\bar{\Phi}_\mathrm{s}$ with respect to the value of the uniform case, $\bar{\Phi}$, as derived in Eq.~(\ref{eqn:Phi}), is given by $\mathcal{E}_\Phi = \bar{\Phi}_\mathrm{s}/\bar{\Phi}-1$. The latter is plotted in Fig.~\ref{fig:plotErrs} in terms of $\alpha_0$ and $\beta_0$, as well as $\lambda$; see also Fig.~\ref{fig:figure_2}(e).
	
	\section{RanGEF as a nuclear cargo}
	
	The nucleotide exchange rate depends on the interaction of RanGTP to RanGEF. The only known nucleotide exchange factor for Ran is bound RCC1 to chromatin. Interestingly, RCC1 is a NLS cargo~\cite{Furuta2016}, which is transported by the Ran cycle.~This leads to a positive feedback that transports RanGEF to regions where RanGTP is located which in turn replenishes that region with even more RanGTP. By assuming that the binding of RanGEF to chromatin is fast, the local rate $\alpha(\boldsymbol{r},t)$ depends on concentration of RanGEF, which we denote by $F(\boldsymbol{r},t)$. We write the nucleotide exchange rate as follows $\alpha(\boldsymbol{r},t) = \alpha_0 F(\boldsymbol{r},t)/F_0$, where $F_0$ is concentration of RanGEF, when all of the RanGEF molecules in the cell are uniformly distributed in the cell nucleus (with no additional RanGEF molecules transported into the nucleus). RanGEF is transported into the nucleus as a NTR--RanGEF complex, and we denote their cytoplasmic and nuclear concentrations by $K_c$ and $K_n$, respectively. In this model, we need to augment the rate $\beta$ with the dissociation of NTR--RanGEF; see Eq.~(13).\\
	\indent
	First, we assume an initial distribution of RanGEF which is uniform, containing only 10\% of the RanGEF molecules in the cell, with the remaining being in the form of cytoplasmic NTR--RanGEF. By solving numerically the system of equations in three-dimensions~\cite{Burns2020}, We make use of a fast spectral method with domain discretization in spherical coordinates and a
	Runge-Kutta scheme for time-stepping~\cite{Burns2020}.	Fig.~\ref{fig:plot_phi_vs_lambda} show the time evolution of $\Phi$, and its dependence on the nuclear pore permeability $\lambda$. 
	The long--time spatial distribution of bound RanGEF is shown in Fig.~\ref{fig:figure_3}(b), which shows radial profiles decaying away from the NE with permeation lengths that decrease with increasing $\gamma$.\\
	\indent
	Second, we assume the initial distribution of RanGEF is given by a various three-dimensional Gaussian clusters of equal standard deviation, being randomly placed within the nucleus; see inset in Fig.~\ref{fig:plot_F_init_vs_F_final} for an example of initial concentration of RanGEF. We simulate 100 different such initial random realizations, as shown in Fig.~\ref{fig:plot_F_init_vs_F_final}.  The long-time spherical angular average of the RanGEF distribution is also shown in Fig.~\ref{fig:plot_F_init_vs_F_final}, which displays a sharp accumulation at the NE despite the initial random distribution of bound RanGEF.\\
	\indent
	Lastly, Fig.~\ref{fig:grid-plots-abi} shows the angular average of nuclear concentrations of nuclear RanGDP, RanGTP, and NTR--RanGTP as function of time and radial distance for the two different initial RanGEF profiles: uniform in Fig.~\ref{fig:grid-plots-abi}(a); and random clusters in Fig.~\ref{fig:grid-plots-abi}(b), with the initial distribution of RanGEF being the configuration shown in the inset plot of Fig.~\ref{fig:plot_F_init_vs_F_final}.
	
	
	%
	
\end{document}